\documentclass[letterpaper,english,aps,prd,twocolumn,groupedaddress]{revtex4-1}
\usepackage[T1]{fontenc}
\usepackage[latin9]{inputenc}
\usepackage{amsmath}
\usepackage{amssymb}
\usepackage{graphicx}

\makeatletter

\usepackage{amsfonts}
\usepackage{bm}

\renewcommand{\vec}[1]{\bm{#1}}
\newcommand{\ceiling}[1]{\left\lceil#1\right\rceil}

\newcommand{\im}{\mathrm{Im}}
\newcommand{\average}[1]{\ensuremath{\left\langle#1\right\rangle}}

\makeatother

\usepackage{babel}
\begin{document}

\preprint{APS/123-QED}

\title{Linear stability analysis of collective neutrino oscillations without
spurious modes}

\author{Taiki Morinaga}

\affiliation{Graduate School of Advanced Science and Engineering, Waseda University,
3-4-1 Okubo, Shinjuku, Tokyo 169-8555, Japan}

\author{Shoichi Yamada}

\affiliation{Advanced Research Institute for Science and Engineering, Waseda University,
3-4-1 Okubo, Shinjuku, Tokyo 169-8555, Japan}

\date{\today}
\begin{abstract}
Collective neutrino oscillations are induced by the presence of neutrinos
themselves. As such they are intrinsically nonlinear phenomena and
are much more complex than linear counterparts such as the vacuum
or MSW oscillations. They obey integro-differential equations, numerical
solutions of which are also very challenging. If one focuses on the
onset of the collective oscillations, on the other hand, the equations
can be linearized and the technique of linear analysis can be employed.
Unfortunately, however, it is well known that such an analysis, when
applied with discretizations of continuous angular distributions,
suffers from the appearance of so-called spurious modes, unphysical
eigenmodes of the discretized linear equations. In this paper, we
analyze in detail the origin of these unphysical modes and present
a simple solution to this annoying problem. We have found that the
spurious modes originate from the artificial production of pole singularities
instead of a branch cut in the Riemann surface by the discretizations.
The branching point singularities in the Riemann surface for the original
undiscretized equations can be recovered by approximating the angular
distributions with polynomials and then performing the integrals analytically.
We demonstrate for some examples that this simple prescription removes
the spurious modes indeed. We also propose an even simpler method:
a piecewise linear approximation to the angular distribution. It is
shown that the same methodology is applicable to the multi-energy
case as well as to the dispersion relation approach that was proposed
very recently. 
\end{abstract}
\maketitle

\section{Introduction}

Collective flavor oscillations of neutrino, in which flavor conversions
are induced by the presence of other neutrinos, are attracting much
interest anew recently, since they may occur near the proto-neutron
star (PNS) surface and affect the dynamics of core-collapse supernova
(CCSN) in a crucial way \cite{PANTALEONE1992128,PhysRevD.48.1462,PhysRevD.72.045003,PhysRevD.74.105014,PhysRevD.74.123004,PhysRevD.84.065008,PhysRevD.85.065008,PhysRevD.87.113010,CHAKRABORTY2016366,1475-7516-2016-03-042,10.1393/ncr/i2016-10120-8,Ann.Rev.Nucl.Part.Sci.66.341,1475-7516-2017-02-019,0004-637X-839-2-132}.
In fact, neutrinos are supposed to play important roles in the mechanism
of CCSN explosions because the gravitational collapse of cores in
massive stars leads to the formation of a shock wave by core bounce,
which is expected to expel the outer part of the star and produce
an explosion as we know it observationally but is stalled in the core
instead; neutrinos emitted copiously from PNS are most likely to be
the instigator of shock revival, which will eventually give rise to
the explosion, by heating up matter on the down-stream side of the
stagnated shock wave. In this so-called neutrino heating scenario,
it is of course the efficiency of the neutrino heating that matters
most. Since the cross sections of absorptions of $\nu_{e}$ and $\bar{\nu}_{e}$,
the main reactions responsible for the heating, are energy-dependent,
$\sigma\propto\epsilon_{\nu}^{2}$, it is crucially important in discussing
the success or failure of the scenario to evaluate the energy spectra
of neutrinos accurately. It is also noted that the energy spectra
are different among neutrino flavors, with $\nu_{e}$ and $\nu_{x}$
(denoting $\nu_{\mu/\tau}$ and $\bar{\nu}_{\mu/\tau}$ collectively)
having the lowest and highest average energies, respectively. It is
then expected that the heating will be enhanced if the flavor conversion
occurs and the energy spectra are swapped between the electron-type
neutrinos with lower energies and other types of neutrinos with higher
ones. 

Matter is dense in the supernova core, in particular, near the PNS
and the neutrino oscillation would be suppressed if there were no
contribution from self-interactions. It was Sawyer \cite{PhysRevD.72.045003,PhysRevD.79.105003,PhysRevLett.116.081101}
who first pointed out the possibility that fast pair-wise flavor conversions
could occur via this collective effect close to the neutrino sphere,
which is the imaginary surface located slightly outside the PNS, from
which neutrinos are effectively emitted, if $\nu_{e}$ and $\bar{\nu}_{e}$
have substantially different angular distributions. If true, neutrinos
of different flavors exchange their energy spectra before they reach
the heating region located at larger distances from the center and
the dynamics of shock revival may be affected as explained above. 

The collective neutrino oscillation is an intrinsically nonlinear
problem, since the potential that induces the oscillation depends
on the consequence of the oscillation itself. As a result, various
interesting phenomena have been demonstrated in the literature \cite{PhysRevD.74.123004,PhysRevD.74.105010,PhysRevD.75.083002,PhysRevD.76.125008,PhysRevD.78.125015,PhysRevLett.103.051105,doi:10.1146/annurev.nucl.012809.104524,PhysRevD.81.073004,PhysRevD.88.045031}
in simplified settings. Much effort has also been put into more realistic
treatments with kinetic equations \cite{PhysRevD.75.125005,PhysRevD.76.125018,PhysRevD.76.085013,PhysRevD.78.085012,0954-3899-36-11-113201,PhysRevD.85.113007,PhysRevD.91.025001}.
On the other hand, the linear stability analysis is conveniently employed
these days to explore conditions, under which the collective oscillation
occurs \cite{PhysRevD.84.053013,PhysRevD.85.113002,PhysRevD.86.085010,PhysRevLett.108.061101,PhysRevLett.108.231102}.
The idea is that neutrinos are initially produced in one of flavor
eigenstates and hence the flavor mixing is treated perturbatively
at least at the beginning of the conversion. It is true that we do
not know from such an analysis what happens once the conversion occurs
and grows to a large mixing of flavors, but it is still important
to know its trigger. This may be particularly the case for CCSN simulations,
since even the most realistic computations have not taken them into
account yet. Hence we had better explore first when and where the
collective oscillations are likely to take place, based on the results
obtained without them. It would be even better if we could implement
a subroutine in the simulation code that conducts such a linear analysis
in real time. 

In the linear analysis, we first linearize the system of equations
that describe the flavor conversion. As a common practice we assume
harmonic oscillations either in time or space or both. In the latter
two cases we further assume that the characteristic wave length of
the flavor oscillation is much shorter than the scale heights of matter
and neutrino densities, i.e., the local approximation is employed.
Then the problem comes down to an eigenvalue problem, in which an
occurrence of complex frequencies or wave numbers means an exponential
growth either in time or in space, respectively, of the flavor conversion\cite{PhysRevD.96.043016}.
Although one may think that solving the linearized equations and finding
eigenvalues and eigenvectors are a done deal, that is not the case.
As a matter of fact, one comes across spurious modes more often than
not, which originate from unavoidable numerical solutions of the equations
and have nothing to do with the physics of our interest \cite{PhysRevD.86.125020}. 

The appearance of such non-physical eigenmodes is easily understood
as follows: the linearized equations are integro-differential equations;
we normally approximate them by discretizing the derivatives and integrals;
then the resultant equations are a linear algebraic system, the dimension
of which is simply determined by how many points are deployed in the
discretization, an arbitrary number as long as it is large enough
to guarantee a certain accuracy. The number of eigenmodes depends
on it, however, being normally equal to it if one takes into account
degeneracy appropriately. Of course the number of the true eigenmodes
should not depend on such an arbitrary number. This implies simply
that not all eigenmodes in the approximation are true. In fact, most
of them are spurious if the dimension of the approximate system is
large. This is ironic, since one deploys a large number of points
for numerical accuracy in the first place but obtains many wrong solutions
instead. It is a solace, however, that if the number is sufficiently
large, the parameter regions, in which the spurious modes emerge,
may not overlap with those of the true modes and we may be able to
distinguish the former from the latter in principle. Note, however,
that we do not know a priori how many points are needed. Too many
of them are just inefficient and certainly bad if one wants to conduct
the real-time survey. 

In this paper we have investigated why these spurious modes appear
more in detail and come up with a simple way to avoid this annoying
problem. The idea is that we perform the integrals in the original
integro-differential equations analytically not by discretizing the
equations but by fitting the distribution functions of neutrinos in
the integrand with appropriate functions such as polynomials, which
allow easy and analytical integrations. In our method, no spurious
mode is produced and all solutions tend to the true ones. The accuracy
of the numerical solutions so obtained solely depends on that of the
fitting. 

Following Sarikas \textit{et al}. \cite{PhysRevD.86.125020}, we start
our exposition in this paper with time-independent perturbations propagating
radially in spherically symmetric backgrounds, in which case the angular
distribution of neutrino in momentum space is axisymmetric with respect
to the local radial direction. We will demonstrate later, however,
that the same method can be applied to more generic cases: non-axisymmetric
perturbations in momentum space \cite{PhysRevD.88.073004,PhysRevLett.111.091101,PhysRevD.89.093001,PhysRevD.90.033004,PhysRevD.92.021702}
as well as arbitrary energy spectra are handled. These days, researchers
in this field are paying their attentions to modes with non-vanishing
frequencies and/or non-radial wave vectors, which can be accommodated
conveniently in the dispersion relations \cite{PhysRevD.92.125030,DUAN2015139,ABBAR201543,1475-7516-2016-04-043,1475-7516-2016-03-042,1475-7516-2017-02-019,PhysRevLett.118.021101}.
It should be stressed that all these cases suffer from the spurious
modes just in the same way. Our method can be also employed in this
new formulation with no difficulty. We will demonstrate finally that
it can be successfully applied to numerical data given on discrete
grid points. In so doing, we use a result of a realistic radiation-hydrodynamic
simulation of CCSN under the assumption of spherical symmetry in space
\cite{0067-0049-214-2-16,0067-0049-229-2-42}. Very recently, our
group has succeeded in first-principles simulations of CCSNe, in which
Boltzmann equations for neutrino transport are solved numerically
under axisymmetry in space without artificial approximations other
than mandatory discretizations. Note that the angular distribution
of neutrino in momentum space has no longer any symmetry. The linear
analysis for such non-axisymmetric angular distributions in background
is currently being undertaken with the same methodology and will be
reported elsewhere.

\section{Linear Stability Analysis}

\subsection{Equations of Flavor Oscillations}

\begin{figure}[tbh]
\includegraphics[width=7cm]{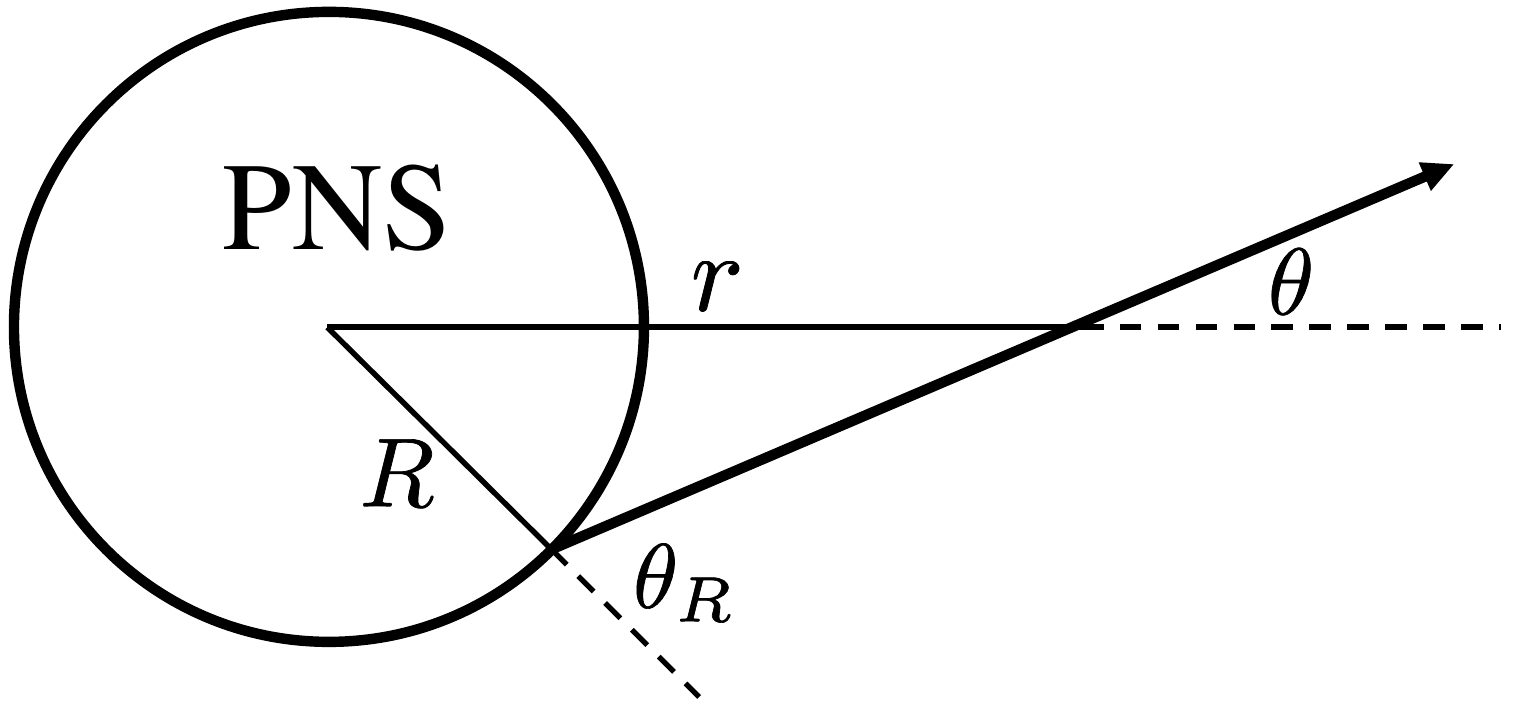} \caption{\label{bulbmodel} A schematic picture of neutrino emissions from
the neutrino sphere. The circle indicates the neutrino sphere, which
is sitting slightly outside the proto neutron star (PNS). Its radius
is $R$. The thick straight line is one of the trajectories of neutrinos
emitted from a point on the neutrino sphere. The emission angle is
denoted by $\theta_{R}$ and is defined as an angle between the trajectory
and the radial direction at the emission point while $\theta$ is
given at each point on the trajectory as displayed in the figure.}
\end{figure}
We begin our discussions with the setup employed in Ref. \cite{PhysRevD.86.125020},
in which neutrinos are emitted semi-isotropically, i.e., generated
uniformly in the outward hemisphere, from each point on the neutrino
sphere in the supernova core (see Fig. \ref{bulbmodel}). They assumed
spherical symmetry for the matter distribution and worked in the two-flavor
oscillation scheme between electron-type and other types (collectively
denoted by $x$ in the following) of neutrinos, which we will follow
here. Then the flavor state of neutrinos can be conveniently described
by $2\times2$ density matrices $\mathbf{\Phi}_{E,u}$, in which the
diagonal components represent the fluxes of individual flavors of
neutrinos that have an energy $E$ and propagate in a direction specified
by $u$; $u$ is defined as $u\equiv\sin^{2}\theta_{R}$ with the
emission angle $\theta_{R}$ measured from the radial direction on
the neutrino sphere of a radius $R$; that off-diagonal components
express the transitions from one flavor to another.

The time evolutions of the density matrices are described by the von
Neumann equations: 
\begin{align}
i\partial_{r}\mathbf{\Phi}_{E,u}=[\mathbf{H}_{E,u},\mathbf{\Phi}_{E,u}],\label{EOM}
\end{align}
where the Hamiltonian matrices are expressed as 
\begin{align}
\mathbf{H}_{E,u,r}= & \left(\dfrac{\mathbf{M}^{2}}{2E}+\sqrt{2}G_{F}\mathbf{N}_{l}\right)\dfrac{1}{v_{u}}\nonumber \\
 & +\dfrac{\sqrt{2}G_{F}}{4\pi r^{2}}\int_{-\infty}^{\infty}dE'\int_{0}^{1}du'\left(\dfrac{1}{v_{u'}v_{u}}-1\right)\mathbf{\Phi}_{E',u'}.\label{Hamiltonian}
\end{align}
In the above equations, $\mathbf{M}^{2}$ is the mass-square matrix,
which describes the vacuum oscillation; the diagonal matrix $\mathbf{N}_{l}=\mathrm{diag}(n_{e}-n_{e^{+}},0)$
expressed with the number densities of electron $n_{e}$ and positron
$n_{e^{+}}$ represents the matter-induced MSW oscillation; it is
tacitly assumed here that other charged leptons do not exist in the
supernova core; $r$ is the distance from the center of proto neutron
star (PNS) and $v_{u}=\cos\theta=\sqrt{1-uR^{2}/r^{2}}$ corresponds
to the radial velocity of neutrino. Note that these equations are
nonlinear actually, since the Hamiltonians include $\mathbf{\Phi}$'s
themselves (the last term in Eq. (\ref{Hamiltonian})). 

We then recast $\mathbf{\Phi}_{\omega,u}$ into
\begin{align}
\mathbf{\Phi}_{\omega,u}=\dfrac{\mathrm{Tr}\mathbf{\Phi}_{\omega,u}}{2}I+\dfrac{F_{\omega,u}^{e}-F_{\omega,u}^{x}}{2}\begin{pmatrix}s_{\omega,u} & S_{\omega,u}\\
S_{\omega,u}^{*} & -s_{\omega,u}
\end{pmatrix},\label{decomposePhi}
\end{align}
where $F_{\omega,u}^{e}$ and $F_{\omega,u}^{x}$ are the fluxes of
$\nu_{e}$ and $\nu_{x}$ at the neutrino sphere, respectively, and
we use $\omega\equiv\varDelta m^{2}/2E$ instead of $E$ to specify
the energy of neutrino just for later convenience. As mentioned above,
it is the non-vanishing off-diagonal components $S_{\omega,u}$ that
indicate the flavor oscillations. They are small compared with the
diagonal components at the beginning of the conversion, since neutrinos
are produced in one of the flavor eigenstates initially. This fact
is the basis of the linear analysis in the following sections. 

\subsection{Linear Stability Condition}

As mentioned just now, since neutrinos are produced in one of the
flavor eigenstates, which corresponds to $s_{\omega,u}=1$ and $S_{\omega,u}=0$
or a diagonal form of $\vec{\Phi}_{E,u}$ in Eq. (\ref{decomposePhi}),
one can assume in the initial phase of the flavor oscillation that
the off-diagonal component $S$ of $\mathbf{\Phi}$ is still small
and can be treated linearly. This is the idea here. Note that the
flavor eigenstate is a fixed point of Eq. (\ref{EOM}) if one ignores
a small off-diagonal component of $\mathbf{M}^{2}$ and the flavor
conversion can be regarded as the instability of this fixed point.
Assuming $|S|\ll1$ and $s=1$, we obtain the linearized equation
for $S$ from Eq. (\ref{EOM}) as follows: 
\begin{align}
i\partial_{r}S_{\omega,u}= & (\omega+u\bar{\lambda})S_{\omega,u}\nonumber \\
 & -\mu\int_{0}^{1}du'\int_{-\infty}^{\infty}d\omega'(u+u')g_{\omega',u'}S_{\omega',u'},\label{linearizedEOM}
\end{align}
in which $g_{\omega,u}$ is the energy spectrum of neutrino and $\lambda=\sqrt{2}G_{F}[n_{e}(r)-n_{\bar{e}}(r)](R^{2}/2r^{2})$
and $\mu=\{\sqrt{2}G_{F}[F_{\bar{\nu}_{e}}(R)-F_{\bar{\nu}_{x}}(R)]/4\pi r^{2}\}(R^{2}/2r^{2})$
correspond to the potentials produced by matter and neutrinos, respectively,
and $\bar{\lambda}$ is defined as $\bar{\lambda}\equiv\lambda+\epsilon\mu$.
Note that the energy spectrum $g$ is normalized as $\int_{-\infty}^{0}d\omega\int_{0}^{1}dug_{\omega,u}=-1$
and hence $\epsilon\equiv\int_{-\infty}^{\infty}d\omega\int_{0}^{1}dug_{\omega,u}$
represents the asymmetry between neutrino and anti-neutrino. 

Considering short-wavelength perturbations, we assume a following
form of solutions: 
\begin{align}
S_{\omega,u}=Q_{\omega,u}e^{-i\Omega r}.\label{PlaneWave}
\end{align}
This is nothing but a (local) normal mode analysis in the spatial
regime. It is valid as long as the wavelength $\sim$ ($1/\Omega$)
is much shorter than the typical length scale in the background configuration.
Inserting Eq. (\ref{PlaneWave}) into Eq. (\ref{linearizedEOM}),
we obtain eigenvalue equations as 
\begin{align}
(\omega+u\bar{\lambda}-\Omega)Q_{\omega,u}=\mu\int_{0}^{1}du'\int_{-\infty}^{\infty}d\omega'(u+u')g_{\omega',u'}Q_{\omega',u'}.\label{EigenEq}
\end{align}
These integral equations have non-trivial solutions for $Q$ only
when $\Omega$ takes one of the eigenvalues. If the eigenvalue has
a positive imaginary part, the corresponding $S$ will grow exponentially
with $r$ at least locally, which implies that the fixed point is
linearly unstable. Since the eigenvalue equations are real, the eigenvalues
are either real numbers or pairs of complex numbers that are conjugate
to each other. It follows then that if there is a non-real eigenvalue,
it immediately means instability. 

Equation (\ref{EigenEq}) can be solved as follows. Since the right
hand side of Eq. (\ref{EigenEq}) is just linear in $u$, $Q_{\omega,u}$
should be expressed as 
\begin{align}
Q_{\omega,u}=\dfrac{a+bu}{\omega+u\bar{\lambda}-\Omega},
\end{align}
with $a$ and $b$ being constants to be determined. Putting this
back into Eq. (\ref{EigenEq}), we obtain the following homogeneous
linear equations for the constants:
\begin{align}
\begin{pmatrix}I_{1}-1 & I_{2}\\
I_{0} & I_{1}-1
\end{pmatrix}\begin{pmatrix}a\\
b
\end{pmatrix}=0.\label{reducedEigenEq}
\end{align}
The elements of the matrix in the above equations are given as the
following integrals: 
\begin{align}
I_{n}\equiv\mu\int_{0}^{1}du\int_{-\infty}^{\infty}d\omega\dfrac{u^{n}g_{\omega,u}}{\omega+u\overline{\lambda}-\Omega}.\label{I}
\end{align}
Finally we obtain the equation to determine the eigenvalue $\Omega$
from the condition that Eq. (\ref{reducedEigenEq}) should have non-trivial
solutions: 
\begin{align}
D(\Omega)\equiv(I_{1}-1)^{2}-I_{0}I_{2}=0.\label{D}
\end{align}

\section{Spurious Modes}

\subsection{Discretization of the Eigenvalue Equations}

We now demonstrate that spurious modes appear when the eigenvalue
equations (\ref{EigenEq}) are solved approximately by discretizing
$Q$ as follows: 
\begin{align}
(\omega_{k}+u_{c}\bar{\lambda}-\Omega)Q_{k,c}=\mu\sum_{i=1}^{N_{\omega}}\sum_{b=1}^{N_{a}}\varDelta\omega\varDelta u(u_{c}+u_{b})g_{i,b}Q_{i,b}.\label{discretizedEigenEq}
\end{align}
Here the integrals in the original equations are replaced with finite
summations, in which $N_{\omega}$ and $N_{a}$ are the numbers of
the bins in the energy- and angle-distributions of neutrinos. %
\mbox{%
Equation (\ref{discretizedEigenEq})%
} is ($N_{\omega}\times N_{a}$) dimensional, homogeneous, linear equations,
which give non-trivial solutions to $Q$ when $\Omega$ takes appropriate
values, which are nothing but the eigenvalues of the matrix derived
from Eq. (\ref{discretizedEigenEq}):
\begin{align}
M_{kc,ib}=(\omega_{k}+u_{c}\bar{\lambda})\delta_{ki}\delta_{cb}-\mu\varDelta\omega\varDelta u(u_{c}+u_{b})g_{i,b}.
\end{align}
 Then $Q$'s are the corresponding eigenvectors.

One may solve Eq. (\ref{D}) instead by evaluating the integrals in
$I_{n}$ numerically. This can be done also by discretizing the integrand
and replacing the integrals with finite sums as 
\begin{align}
I_{n}=\mu\sum_{i=1}^{N_{\omega}}\sum_{b=1}^{N_{a}}\varDelta\omega\varDelta u\dfrac{u_{b}^{n}g_{i,b}}{\omega_{i}+u_{b}\overline{\lambda}-\Omega}.\label{discretizedIn}
\end{align}
We note that the two methods are essentially the same because they
are both obtained by approximating the distribution function $g_{\omega,u}$
as 
\begin{align}
g_{\omega,u}=\sum_{i=1}^{N_{\omega}}\sum_{b=1}^{N_{a}}g_{\omega_{i},u_{b}}\varDelta\omega\varDelta u\delta(\omega-\omega_{i})\delta(u-u_{b}).\label{discretizedg}
\end{align}
The same eigenvalues and eigenvectors are hence obtained in both approaches.
What is important here is that all of these supposedly approximate
solutions do not actually correspond to the real solutions of the
original equations. This is understood from the fact that the number
of the solutions for the former depends on the number of bins employed.
The solutions of Eq. (\ref{discretizedEigenEq}) that do not correspond
to any real modes are called spurious modes.

\subsection{Characteristics of Spurious Modes}

We elucidate the features of the spurious modes, using a simple model
with a monochromatic energy distribution \cite{PhysRevD.86.125020}:
\begin{align}
g_{\omega,u}=\left[(1+\epsilon)\delta(\omega-\omega_{0})-\delta(\omega+\omega_{0})\right]B(u).\label{monochromaticg}
\end{align}
In this expression, $B(u)$ is an angular distribution and is assumed
in this section to be 
\begin{align}
B(u)=1,
\end{align}
which corresponds to the semi-isotropic emission from the neutrino
sphere. For this neutrino distribution in the background, we can perform
the integrals in $I_{n}$ analytically as 
\begin{align}
I_{n}(\Omega)=\dfrac{\mu}{\bar{\lambda}}\left[(1+\epsilon)\left\{ U_{+}^{n}\ln\left(1-\dfrac{1}{U_{+}}\right)+\sum_{j=0}^{n-1}\dfrac{1}{n-j}U_{+}^{j}\right\} \right.\nonumber \\
\left.-\left\{ U_{-}^{n}\ln\left(1-\dfrac{1}{U_{-}}\right)+\sum_{j=0}^{n-1}\dfrac{1}{n-j}U_{-}^{j}\right\} \right],\label{integratedIn}
\end{align}
where we define 
\begin{align}
U_{+}(\Omega)\equiv\dfrac{\Omega-\omega_{0}}{\overline{\lambda}},\ \ U_{-}(\Omega)\equiv\dfrac{\Omega+\omega_{0}}{\overline{\lambda}}.
\end{align}
Then Eq. (\ref{D}) can be solved numerically to a desired accuracy
without difficulties. We hence regard the modes so obtained as true
modes and use them as reference in the following. 
\begin{figure}[!tbh]
\includegraphics[width=7cm]{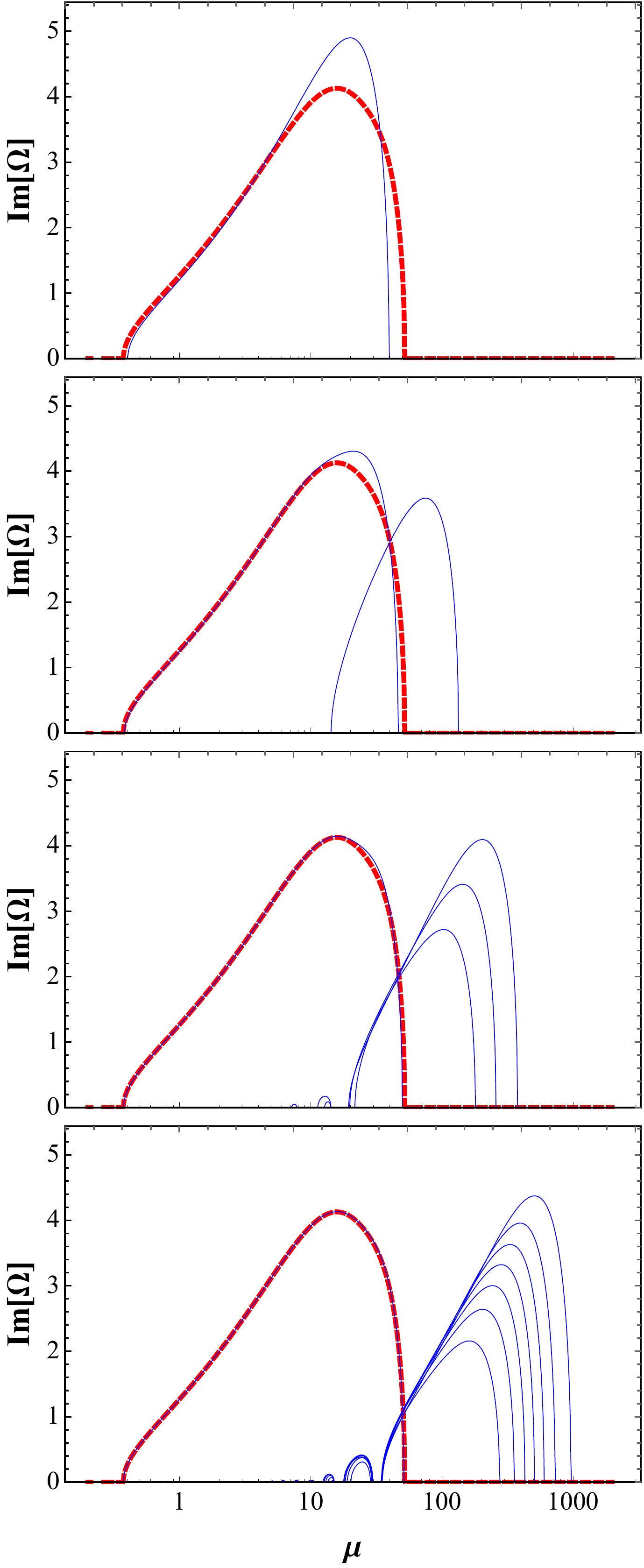} \caption{\label{kappa-mu_b1bins} The imaginary part of $\Omega$, $\im\Omega$,
as a function of $\mu$ for the exact (red dashed lines) and approximate
(blue solid lines) solutions for different numbers of angular bins.
From top to bottom the numbers of bins $N_{a}$ are 1, 2, 4 and 8.
We employ $B(u)=1$ for these calculations. Note that only the leftmost
branch of the approximate solutions (blue solid lines) approaches
the true solutions (red dashed lines) as $N_{a}$ increases and is
regarded as the (approximate) physical solution whereas all the other
approximate solutions are spurious, moving rightwards away from the
true solutions as $N_{a}$ increases.}
\end{figure}
\begin{figure}[!tbh]
\includegraphics[width=7cm]{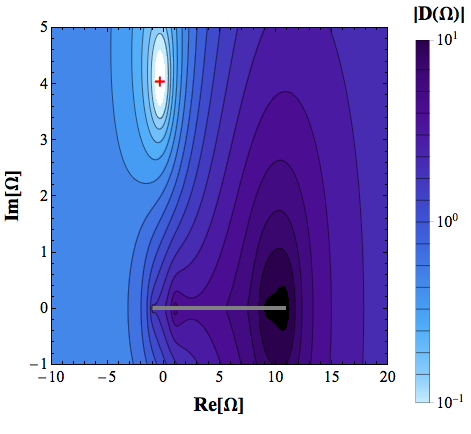} \caption{\label{contour_b1exact}Absolute values of $D(\Omega)$ on the complex
plain of $\Omega$. The integrals in $I_{n}$ are performed analytically.
The plus indicates one of the complex zero points of $D(\Omega)$.
The gray line is the branch cut of $D(\Omega)$ on the Riemann surface.
Note that only one of the zero points approaches the true one as $N_{a}$
increases.}
\end{figure}
\begin{figure*}[htb]
\centering \includegraphics[width=7cm]{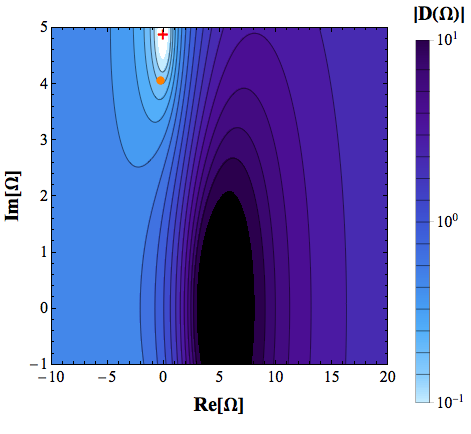} \includegraphics[width=7cm]{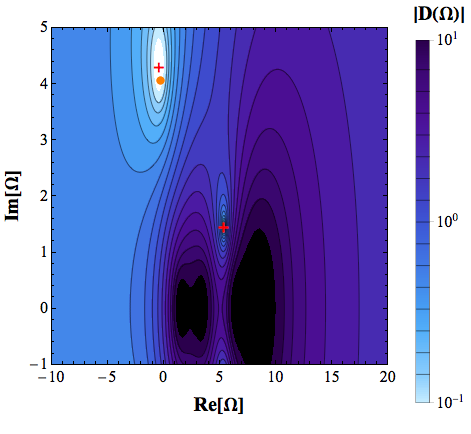}
\includegraphics[width=7cm]{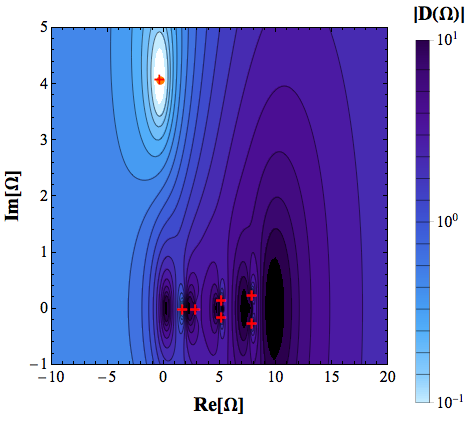} \includegraphics[width=7cm]{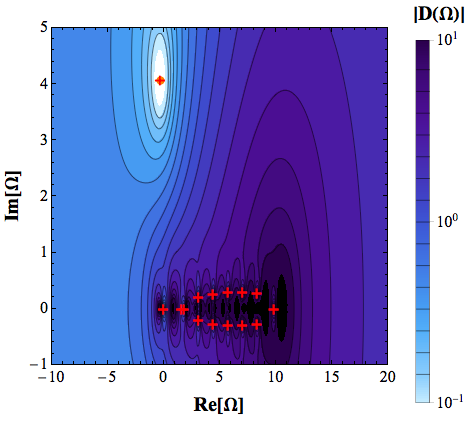}
\caption{\label{contour_b1bins} Same as Fig. \ref{contour_b1exact} except
that the integrals are approximately evaluated with the discretization.
The numbers of angular bins $N_{a}$ are $1$, $2$, $4$ and $8$
clockwise from top left. Plus signs indicate some of the zero points
of $D(\Omega)$ whereas orange dots correspond to the exact root of
$D(\Omega)$ given in Fig. \ref{contour_b1exact}.}
\end{figure*}

We find that Eq. (\ref{D}) has 2 solutions or eigenvalues $\Omega$
for $\epsilon=0.5$ and $\lambda=0$. They are either both real or
complex conjugate to each other, depending on the value of $\mu$.
See the dashed line in Fig. \ref{kappa-mu_b1bins}, in which we show
the imaginary parts of the true and approximate solutions as a function
of $\mu$. Note that they are by definition non-vanishing only of
the complex solutions, for which we present only those solutions with
positive imaginary parts, since others are complex conjugate to them.
One can see that there is a range of $\mu$, $\sim0.4<\mu<\sim50$,
in which the true eigenvalue acquires a non-vanishing imaginary part.
On the other hand, the discretized equations (Eq. (\ref{discretizedEigenEq}))
with $N_{a}$ angular bins yield $2N_{a}$ eigenvalues (solid lines
in Fig. \ref{kappa-mu_b1bins}). It is clear that only one of these
solutions approximate the true solution, which is indeed corroborated
in the figure by the fact that it comes closer to the true solution
as the number of angular bins is increased. This is not the case of
other solutions, on the other hand. In fact, they have different ranges
of $\mu$, where they become complex and have non-vanishing $\im\Omega$'s,
and those regions move away from the true one in general as we deploy
more angular bins. They are the spurious modes we are concerned with
in this paper. Their characteristics mentioned above are not our original
findings but just the reproduction of what were presented in \cite{PhysRevD.86.125020}.
We show them here again because we will begin our analysis with these
simple solutions. 

The inspection of $D(\Omega)$ gives us the hint of the reason why
the spurious modes are produced by the discretization. Figure \ref{contour_b1exact}
exhibits in the complex $\Omega$ plane the absolute values of $D(\Omega)$
obtained from the analytical integrations of $I_{n}$ (Eq. (\ref{integratedIn}))
whereas Fig. \ref{contour_b1bins} displays the same quantities but
for $I_{n}$ evaluated approximately with the discretization (Eq.
(\ref{discretizedIn})). When exactly calculated, $D(\Omega)$ has
a branch cut from $-\omega_{0}$ to $\omega_{0}+\bar{\lambda}$ on
the real axis in the Riemann surface. There is a discontinuity in
the imaginary part of $D(\Omega)$ on this cut. In fact, it has the
same absolute value but has opposite signs just above ($\im\Omega>0$)
and below ($\im\Omega<0$) the cut. Note that in Fig. \ref{contour_b1exact}
the cut is indicated with a gray line although $|D(\Omega)|$ is continuous.
In Fig. \ref{contour_b1bins}, on the other hand, not the branch cut
but poles appear on and near the line, at which the cut should be
located, and $D(\Omega)$ is analytic except on these poles. This
feature is unchanged and only the number of poles increases if we
deploy larger numbers of bins. 

This situation can be demonstrated more explicitly in equations. When
we evaluate the integrals in $I_{n}$ by the discretization, they
are expressed as 
\begin{align}
I_{n}=\mu\sum_{b=1}^{N_{a}}u_{b}^{n}B(u_{b})\left(\dfrac{1+\epsilon}{\omega_{0}+u_{b}\overline{\lambda}-\Omega}-\dfrac{1}{-\omega_{0}+u_{b}\overline{\lambda}-\Omega}\right).
\end{align}
These are sums of fractional functions of $\Omega$, the poles of
which are $\omega_{0}+u_{b}\bar{\lambda}$ and $-\omega_{0}+u_{b}\bar{\lambda}$.
This difference in the singularity structures in $D(\Omega)$ is responsible
for the appearance of the spurious modes. This may be understood as
follows. When we search for the roots of $D(\Omega)$, we first combine
the fractions in $I_{n}$ to a single fraction with the common denominator
and then seek for the roots of the numerator, which is a polynomial
in $\Omega$. Its degree becomes larger as the number of bins is increased.
As a result, one obtains more roots inevitable. It should be now apparent
that the point here is that the branch cut is replaced with the poles
by the discretization of $g_{\omega,\mu}$ in Eq. (\ref{discretizedg}).
This leads in fact to $D(\Omega)$ that never approaches the true
one even if one increases the number of bins as long as it is finite.
It was argued in \cite{PhysRevD.86.125020} that if one deploys a
large enough number of bins, the range of $\mu$, where the spurious
modes develop non-vanishing imaginary parts, may not overlap with
those for the true modes and the spurious modes become harmless. As
can be seen in Fig. \ref{kappa-mu_b1bins}, however, the behavior
of the spurious modes is not simple: the range strongly depends on
the number of bins; it happens in fact that all the spurious modes
become real in the $\mu$-range of interest at some $N_{a}$ but the
imaginary part becomes non-vanishing again at larger $N_{a}$. The
problem is hence that we simply do not know a priori what number of
$N_{a}$ is appropriate.

\section{Analytical-Integration Approach}

\subsection{Polynomial Approximation}

The approximate evaluation of the integrals $I_{n}$ with the discretization
of the distribution function generates the poles instead of the branch
cut in $D(\Omega)$ and is the ultimate culprit for the spurious roots
of $D(\Omega)$. It is hence a natural expectation that we can avoid
the spurious modes if the integrals in $I_{n}$ are approximated in
such a way that no poles would be generated. A simple way to do this
is to approximate the distribution function $g_{\omega,u}$ not with
the delta function but with a polynomial. Here we continue to assume
for simplicity that the neutrino distribution is monochromatic and
$g_{\omega,u}$ is expressed as in Eq. (\ref{monochromaticg}). $B(u)$,
on the other hand, is an arbitrary continuous function. 

Let us suppose that $B(u)$ is approximated as a polynomial of $d$-th
degree: 
\begin{align}
B(u)=\sum_{k=0}^{d}b_{k}u^{k}.
\end{align}
Then, using the following formula 
\begin{align}
\int_{0}^{1}du\dfrac{u^{n}}{u-x}=x^{n}\ln\left(1-\dfrac{1}{x}\right)+\sum_{j=0}^{n-1}\dfrac{1}{n-j}x^{j},
\end{align}
we can perform the integrals in $I_{n}$ analytically to obtain 
\begin{align}
I_{n}(\Omega)= & \dfrac{\mu}{\bar{\lambda}}\sum_{k=0}^{d}b_{k}\biggl[\biggl\{(1+\epsilon)U_{+}^{k+n}\ln\left(1-\dfrac{1}{U_{+}}\right)\nonumber \\
 & +\sum_{j=0}^{k+n-1}\dfrac{1}{k+n-j}U_{+}^{j}\biggr\}-\biggl\{ U_{-}^{k+n}\ln\left(1-\dfrac{1}{U_{-}}\right)\nonumber \\
 & +\sum_{j=0}^{k+n-1}\dfrac{1}{k+n-j}U_{-}^{j}\biggr\}\biggr].\nonumber \\
\label{Iintegrated}
\end{align}
Here and henceforth, the summation is defined to be zero when the
upper limit is smaller than the lower limit; the logarithmic function
$\ln(z)$ should take the principal value. It is apparent that in
this approximation, $I_{n}$ consists of polynomial or logarithmic
functions of $\Omega$, which have no pole. 
\begin{figure}[!tbh]
\includegraphics[width=7cm]{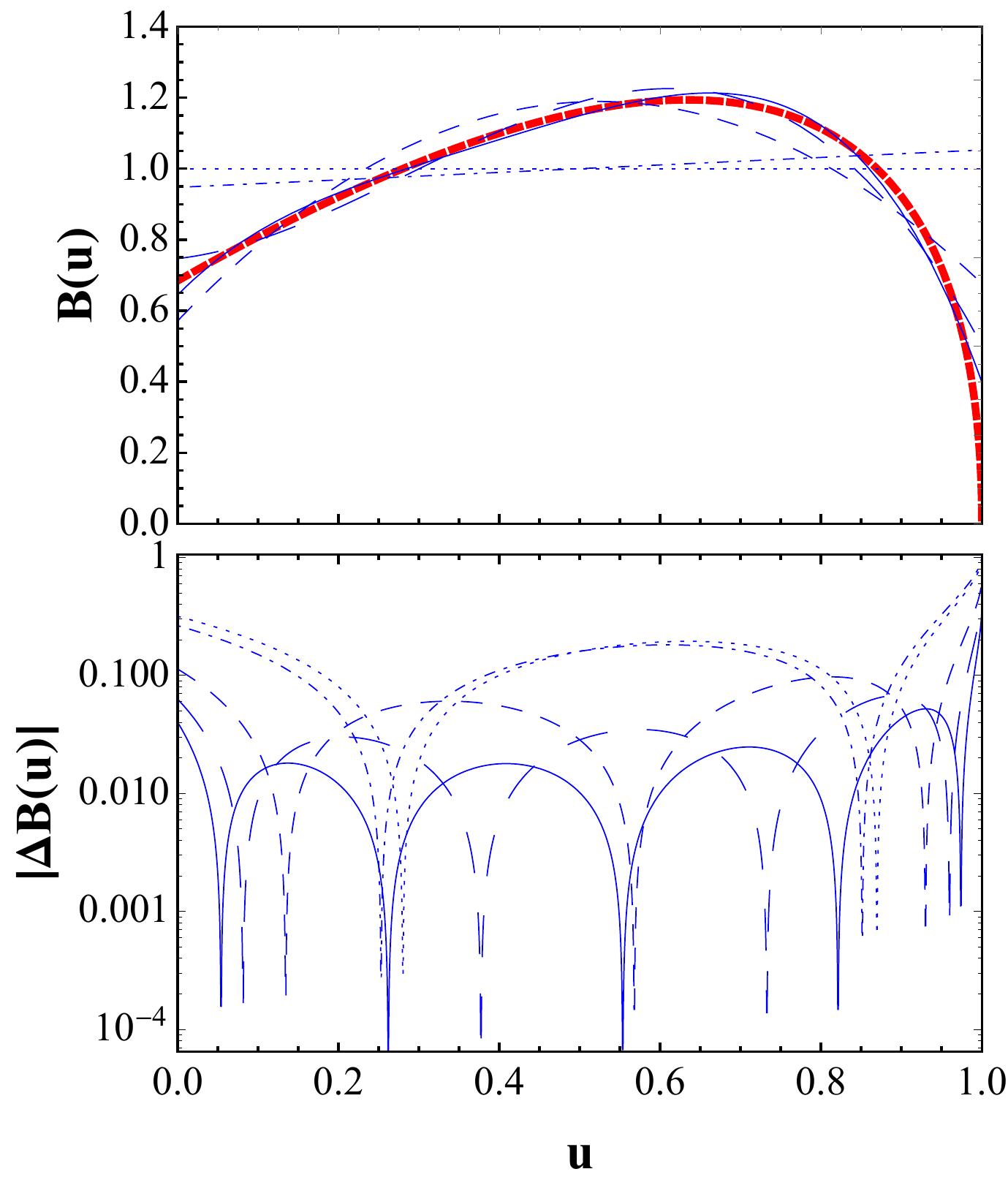} \caption{The original function (red line) and the polynomial approximations
(blue lines) of $B(u)$ (top panel) and the absolute values of the
differences between them (bottom panel). The dotted, dot-dashed, short
dashed, long dashed and solid lines correspond, respectively, to the
polynomial degrees $d$ of $0$, $1$, $2$, $3$ and $4$.}
\label{b-u_btoys} 
\end{figure}
\begin{figure}[!tbh]
\includegraphics[width=7cm]{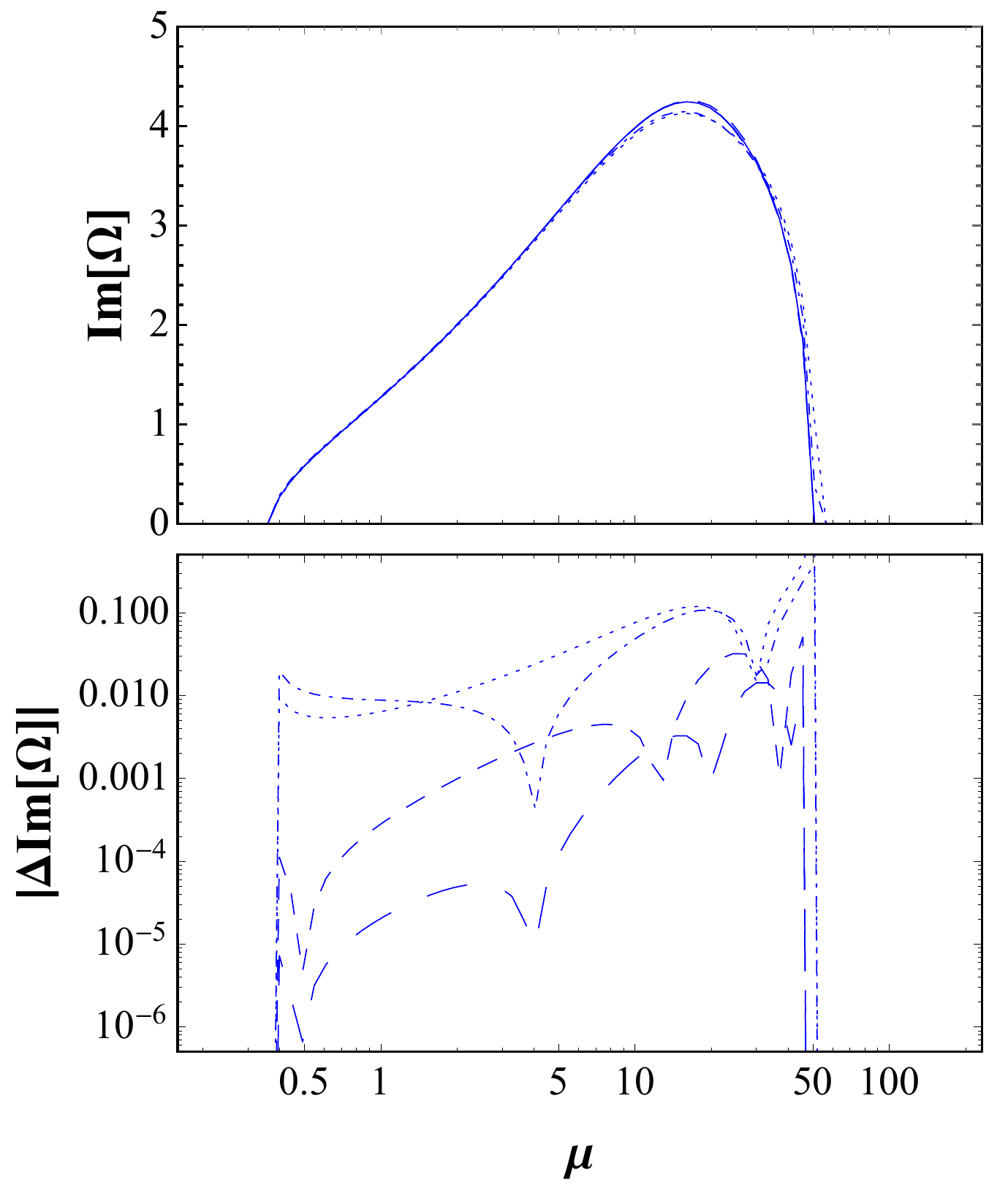} \caption{Top: the imaginary part of $\Omega$ as a function of $\mu$ for the
solutions of Eq. (\ref{D}) with $I_{n}$ given in Eq. (\ref{Iintegrated}).
The line styles denote the polynomial degrees $d$ as in Fig. \ref{b-u_btoys}.
Bottom: the absolute values of the differences between the exact and
approximate solutions for $d=4$. }
\label{kappa-mu_btoys} 
\end{figure}
\begin{figure}[htb]
\includegraphics[width=7cm]{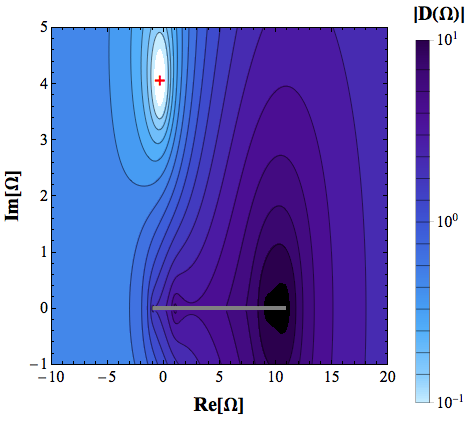} \includegraphics[width=7cm]{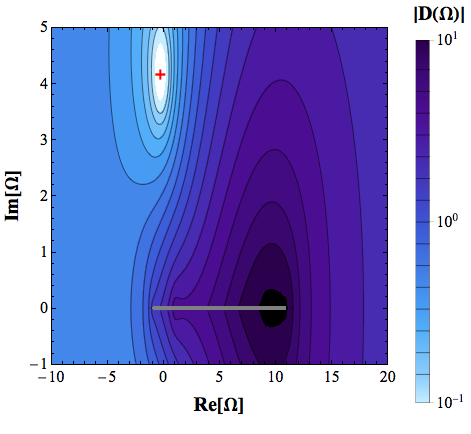}
\caption{The absolute values of $D(\Omega)$ on the complex plain of $\Omega$
for $B(u)$ given in Fig. \ref{b-u_btoys} with $\mu=20$. The degree
of the polynomial is set to $d=1$ and 4 in the top and bottom panels,
respectively.}
\label{contour_btoys} 
\end{figure}

We now demonstrate that this method works as expected indeed. We employ
a simple angular distribution adopted in Ref. \cite{PhysRevLett.108.231102}.
$B(u)=(N_{\bar{\nu}_{e}}U_{\bar{\nu}_{e}}(u)+N_{\bar{\nu}_{x}}U_{\bar{\nu}_{x}}(u))/(N_{\bar{\nu}_{e}}-N_{\bar{\nu}_{x}})$
with $N_{\bar{\nu}_{e}}=1$, $N_{\bar{\nu}_{x}}=0.62$, $U_{\bar{\nu}_{e}}(u)=(3/2)(1-u)^{1/2}$
and $U_{\bar{\nu}_{x}}(u)=2(1-u)$; $\epsilon$ is set to 0.5. Note
that the ratio of $N_{\nu_{e}}$ to $N_{\nu_{x}}$ is the same as
that of $N_{\bar{\nu}_{e}}$ to $N_{\bar{\nu}_{x}}$ in this setup:
$N_{\nu_{e}}:N_{\bar{\nu}_{e}}:N_{\nu_{x}}:N_{\bar{\nu}_{x}}=1.5:1:0.93:0.62$.
This $B(u)$ is approximated with a polynomial function, which is
determined by minimizing the integrated square of errors from the
original function. We show in Fig. \ref{b-u_btoys} both the original
function and the polynomial approximations of $B(u)$ as well as the
errors. As the degree of the polynomials increases, the approximation
gets better just as expected. We then solve Eq. (\ref{D}) with Eq.
(\ref{Iintegrated}). The results are shown Fig. \ref{kappa-mu_btoys}.
It is obvious that no spurious mode is produced in this approach and
the approximate solutions converge to the supposedly exact one as
the degree of polynomial is increased and $B(u)$ is better approximated.
The absence of the spurious modes is also corroborated in Fig. \ref{contour_btoys},
in which we present the absolute values of $D(\Omega)$ obtained approximately
in this method. It is clear that there is only one zero point. 

It is evident from Eq. (\ref{Iintegrated}) that the polynomial approximation
generates a branch cut in the Riemann surface along the real axis
from $\Omega=-\omega_{0}$ to $\omega_{0}+\bar{\lambda}$, the same
feature as for the exact $D(\Omega)$. The important thing in avoiding
the spurious mode are the fact that the integrals in $I_{n}$ can
be done analytically and that the original features of the Riemann
surface are maintained by the approximation for $g_{\omega,u}$. The
use of the polynomial functions is hence not essential and any functions
will be fine as long as they satisfy these conditions. For example,
one may use not only $u^{n}$ but also $u^{n}\sqrt{1-u^{2}}$ to expand
$B(u)$. In fact, it is easily confirmed that the integrals can be
done still analytically and the essential feature of the Riemann surface
is retained also in this case. We have confirmed that no spurious
mode appears then.

The choice of base functions should depend on the distribution function.
It is certainly better if the distribution function is approximated
accurately by a smaller number of the base functions. Note that if
the approximation is not accurate, it may happen that a new spurious
eigenvalue appears and/or a true eigenvalue disappears. It is important
in this context to point out that the number of true modes is related
with that of the \textquotedbl{}crossings\textquotedbl{} in the distribution
function \cite{PhysRevD.84.053013}, i.e., the neutrino oscillation
tends to be triggered when the energy or angular distribution changes
sign. We should hence approximate the distribution function so that
the number of \textquotedbl{}crossings\textquotedbl{} should be unchanged.

\subsection{Piecewise Constant Approximation}

In the previous section we approximated the angular distribution of
neutrino as a whole. We stressed particularly the importance of retaining
the essential feature of the Riemann surface in the approximation.
This may be relaxed, though. As a matter of fact, we demonstrate in
this section that a piecewise constant approximation to the angular
distribution is sufficient to avoid the spurious modes. This approximation
will be of practical use in dealing with numerical data, which are
normally provided only at discrete grid points. We divide the interval
$[0,1]$ into $N_{a}$ sub-intervals, $\{[s_{b-1},s_{b}]\}\ (b=1,2,\cdots,N_{a},\ s_{0}=0,\ s_{N_{a}}=1)$,
and the function $u^{n}B(u)$ in the integrand of $I_{n}$ is approximated
as a constant $u_{b}^{n}B(u_{b})$ in each interval with $u_{b}\equiv(s_{b-1}+s_{b})/2$.
Then the integrals can be performed analytically for each interval
and the results are given as 

\begin{alignat}{2}
I_{n}(\Omega)= &  & \mu\sum_{b=1}^{N_{a}}\int_{s_{b-1}}^{s_{b}}duu_{b}^{n}B(u_{b})\nonumber \\
 &  & \times\biggl(\dfrac{1+\epsilon}{\omega_{0}+u\bar{\lambda}-\Omega} & -\dfrac{1}{-\omega_{0}+u\bar{\lambda}-\Omega}\biggr)\nonumber \\
= &  & \dfrac{\mu}{\bar{\lambda}}\sum_{b=1}^{N_{a}}u_{b}^{n}B(u_{b})\biggl[(1+\epsilon) & \ln\dfrac{\omega_{0}+s_{i}\bar{\lambda}-\Omega}{\omega_{0}+s_{i-1}\bar{\lambda}-\Omega}\nonumber \\
 &  & -\ln & \dfrac{-\omega_{0}+s_{i}\bar{\lambda}-\Omega}{-\omega_{0}+s_{i-1}\bar{\lambda}-\Omega}\biggr]
\end{alignat}
We have confirmed that Eq. (\ref{D}) with this representation of
$I_{n}$ do not produce spurious modes. Note that the essential feature
of the Riemann surface \textit{is} changed in this case. In fact,
there is still the same branch cut along the real axis but it is actually
a union of sub-cuts from $\Omega=\pm\omega_{0}+s_{i-1}\bar{\lambda}$
to $\Omega=\pm\omega_{0}+s_{i}\bar{\lambda}$ and the branching singularities
occur at both ends of each sub-cut. In spite of this change in the
Riemann surface, the spurious modes do not appear.

One may think that the piecewise constant approximation considered
here is equivalent to the discretization, which was responsible for
the generation of the spurious mode. This is not the case, however.
The point is that we have exactly performed the integrals in $I_{n}$
in each sub-interval for the constant angular distribution whereas
in the discretization approximation the integrals are evaluated approximately
by sampling at a finite number of points $u_{i}$ (see Eq. (\ref{discretizedEigenEq})).
In the latter case, even if the eigenvector $Q_{\omega,u}$ diverges
at $u\in[0,1]\backslash\{u_{i}\}$, the approximated integrals are
not affected and remain finite although they would diverge if the
integrals were done exactly for such $Q_{\omega,u}$. They cannot
be an eigenvector in the piecewise constant approximation, either,
since the integrals are divergent. Note also that the discontinuities
at the boundaries of the sub-intervals are responsible for the appearances
of the extra branching points in the piecewise constant approximation;
they pose no serious problem, though, since the integrals approach
the true values as the number of intervals increases; this is in sharp
contrast to the discretization approach.

One should be reminded that the accuracy still matters in the piecewise
constant approximation. Indeed a new spurious mode may appear and/or
a true mode may disappear if the approximation is not very good. One
may hence need to deploy many sub-intervals. One may well consider
to use a piecewise linear (or higher-order) approximation instead
of the piecewise constant one. After all, how many sub-intervals or
what base functions should be used depends on the distribution function
at hand. What is important regardless, however, is that the problem
with the spurious modes arising from the discretization of the distribution
is resolved in principle in the polynomial or piecewise constant approximation.

\subsection{Multi-Energies}

So far we have dealt with a monochromatic distribution. In reality,
however, neutrinos have continuous energy spectra. Then the $\omega$
integration in $I_{n}$ becomes non-trivial. One may think that it
is necessary to approximate $g_{\omega,u}$ so that the double integrals
in $I_{n}$ could be done analytically and should not change the structure
of the Riemann surface. Fortunately, this is not so difficult as it
sounds.

What we need to respect most is to retain the essential feature of
the Riemann surface. One way to do this may be to change the integral
variable $\omega$ to $p\equiv\omega+u\bar{\lambda}$ as 
\begin{align}
I_{n}=\mu\int_{-\infty}^{\infty}dp\dfrac{G_{n}(p)}{p-\Omega}\label{poleintegral}
\end{align}
with 
\begin{align}
G_{n}(p)\equiv\int_{0}^{1}duu^{n}g_{p-u\bar{\lambda},u}.\label{Gnofp}
\end{align}
The latter function $G_{n}(p)$ can be evaluated by the discretization
of the integrand in $u$. We can then approximate it with a polynomial
function and perform the integral in Eq. (\ref{poleintegral}) analytically
just as in the previous sections.

There are some complications in this method, though. Since the integral
range extends to infinity, the integral would be divergent if $G_{n}(p)$
were approximated with polynomial functions nominally. In reality,
$G_{n}(p)$ goes to 0 as $|p|\to\infty$ and the integrals are convergent.
This suggests that $G_{n}(p)$ should be approximated with functions
that ensure the convergence of the integrals. It is normally difficult
to perform the integrals analytically for such kinds of functions,
however, and even if it can be done, the results tend to be complicated.
From a practical point of view, we had better truncate the integral
at range $p_{\mathrm{min}}$ and $p_{\mathrm{max}}$. Then the branch
cut in the Riemann surface is shrunk to the line connecting $p_{\mathrm{min}}$
and $p_{\mathrm{max}}+\bar{\lambda}$ on the real axis although it
is the entire real axis in the exact case. This may not be so serious
a problem, though, since it will affect only those modes with very
small imaginary parts. 

The piecewise approximation discussed in %
\mbox{%
section IV B%
} will be also available in the present case. As a matter of a fact,
$G_{n}(p)$ often looks like a Fermi-Dirac distribution and the polynomial
approximation over the whole interval is not appropriate. 

The method given above may not be suitable after all for the analysis
of numerical data, which provide the values of $g_{\omega,u}$ only
at discrete points $\{(\omega_{i},u_{b})\}$, since it is then difficult
to obtain $G_{n}(p)$ at desirable points of $p$. Fortunately, we
have another (and simpler indeed) option to avoid the spurious modes.
In this method we approximate $g_{\omega,u}$ polynomially in $u$
for a discrete set of $\omega$ just as in the monochromatic case
and then simply take the sum. This is equivalent to approximate $g_{\omega,u}$
as 
\begin{align}
g_{\omega,u}\simeq\sum_{i=1}^{N_{\omega}}\varDelta\omega\delta(\omega-\omega_{i})\sum_{k=0}^{d}b_{i,k}u^{k}.
\end{align}
Then the integrals in $I_{n}$ can be performed to obtain 
\begin{align}
I_{n}(\Omega)=\dfrac{\mu}{\overline{\lambda}}\sum_{i=1}^{N_{\omega}}\sum_{k=0}^{d}\varDelta\omega\ b_{i,k}\biggl\{ U_{i}^{k+n}\ln\left(1-\dfrac{1}{U_{i}}\right)\nonumber \\
+\sum_{j=0}^{k+n-1}\dfrac{1}{k+n-j}U_{i}^{j}\biggr\},\label{multiintegrated}
\end{align}
where $U_{i}$'s are defined as 
\begin{align}
U_{i}(\Omega)\equiv\dfrac{\Omega-\omega_{i}}{\overline{\lambda}}.
\end{align}
Note that the branch cut is again shrunken to a finite interval $[\omega_{1},\omega_{N_{\omega}}+\bar{\lambda}]$
on the real axis and there occurs many branching singularities on
the cut. It turns out that this is sufficient to avoid the spurious
modes. 

\begin{figure}[!tbh]
\includegraphics[width=6.8cm]{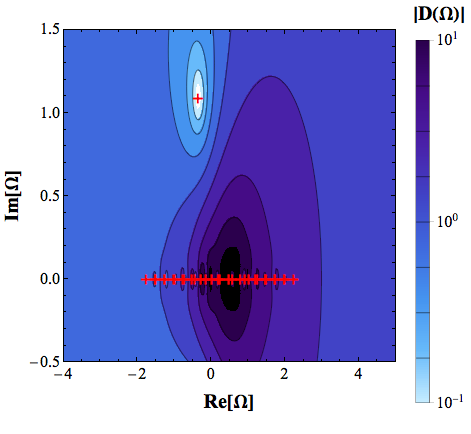} \includegraphics[width=6.8cm]{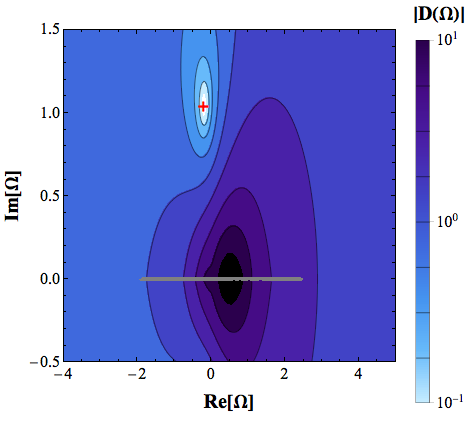}
\includegraphics[width=6.8cm]{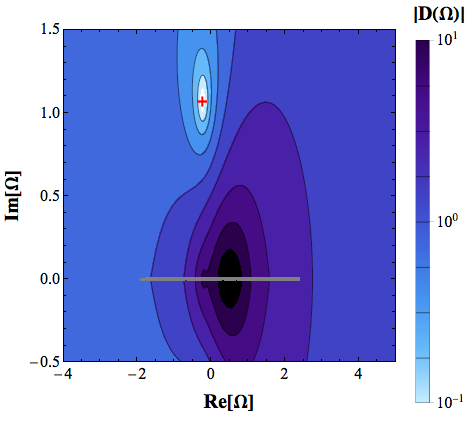} \caption{Absolute values of $D(\Omega)$ on the complex plain of $\Omega$
for the simple multi-energy and multi-angle model given in the text.
$D$ is evaluated in the discretization approximation with $N_{\omega}=16,N_{a}=2$
(top panel) or calculated in Eq. (\ref{poleintegral}) with 16 sub-intervals
in energy and with $G_{n}(p)$ being evaluated in Eq. (\ref{Gnofp})
with 16 angular bins (middle panel) or obtained with Eq. (\ref{multiintegrated})
for $N_{\omega}=32,d=4$ (bottom panel). Zero points are marked with
red pluses and the branch cuts are indicated with the gray lines.}
\label{contour_multiene} 
\end{figure}

Figure \ref{contour_multiene} presents an example. We show the behavior
of $|D(\Omega)|$ for a simple multi-energy and multi-angle distribution
employed in Ref. \cite{PhysRevLett.108.231102}: 
\begin{align}
g_{\omega,u}= & \dfrac{\varDelta m^{2}}{2\omega^{2}}[\theta(\omega)\{F_{\nu_{e}}(E(\omega),u)-F_{\nu_{x}}(E(\omega),u)\}\nonumber \\
 & +\theta(-\omega)\{F_{\bar{\nu}_{e}}(E(\omega),u)-F_{\bar{\nu}_{x}}(E(\omega),u)\}]
\end{align}
with 
\begin{align}
F_{\nu_{\alpha}}=N_{\nu_{\alpha}}\times\varphi_{\nu_{\alpha}}(E)\times\tilde{U}_{\nu_{\alpha}}(u),
\end{align}
\begin{align}
\varphi_{\nu_{\alpha}}(E) & =\dfrac{(1+\alpha)^{1+\alpha}}{\Gamma(1+\alpha)}\dfrac{E^{\alpha}}{\average{E_{\nu}}^{\alpha+1}}\exp\left[-\dfrac{(1+\alpha)E}{\average{E_{\nu}}}\right]
\end{align}
and 
\begin{align}
\tilde{U}_{\nu_{\alpha}}(u) & =\left(\dfrac{\beta_{\alpha}}{2}+1\right)(1-u)^{\beta_{\alpha}/2}.
\end{align}
Here we set the model parameters as follows: $(\average{E_{\nu_{e}}},\average{E_{\bar{\nu}_{e}}},\average{E_{\nu_{x}}})=(12,15,18)\mathrm{MeV}$,
$N_{\nu_{e}}:N_{\bar{\nu}_{e}}:N_{\nu_{x}}=1.5:1:0.62$, $\beta_{e}=1,\beta_{x}=3$,
$\Delta m^{2}=10$, $\lambda=0$ and $\mu=1$. Zero points are marked
with red pluses in the figure. In the top panel, we give the result
for the discretization approximation. It is apparent that a lot of
spurious modes appear on the real axis. On the other hand, there is
no such spurious mode in the middle and the bottom panels, in which
$I_{n}$ is evaluated with Eqs. (\ref{poleintegral}) and (\ref{multiintegrated}),
respectively. We remark that the energy integrals are truncated at
$\omega\approx\pm2$ and, as a result, the branch cut runs approximately
from $\sim-2$ to $\sim3$, which is indicated again by the gray lines
in the figure. 

\section{Applications to the Dispersion Relation Approach}

\subsection{Linear Analysis in the Dispersion Relation Approach}

More recently, Izaguirre et al. \cite{PhysRevLett.118.021101} proposed
an elegant approach to the linear analysis of the collective neutrino
oscillation based on the \textquotedbl{}dispersion relation\textquotedbl{}.
In their approach, not only the instability in the spatial regime
but also in the temporal regime can be handled on the same basis.
In this section we demonstrate that the method proposed in this paper
can be applied also to this dispersion relation approach.

We begin with a brief review of the dispersion relation approach.
The equations of motion for free-streaming neutrinos without collisions
but with flavor conversions are written as \cite{SIGL1993423,PhysRevD.78.085017}
\begin{align}
(\partial_{t}+\vec{v}\cdot\boldsymbol{\nabla}_{\vec{r}})\rho=i[\rho,H],\label{freestreaming}
\end{align}
where the Hamiltonian is given as 
\begin{align}
H=\dfrac{M^{2}}{2E}+v^{\mu}\Lambda_{\mu}\dfrac{\sigma_{3}}{2}+\sqrt{2}G_{F}\int d\Gamma'v^{\mu}v_{\mu}'\rho'.
\end{align}
In these equations $\rho(t,\mathbf{r},\mathbf{p})$ is again the density
matrix, $(v^{\mu})=(1,\vec{v})$ is the neutrino four velocity and
$\Lambda^{\mu}$ consists of $\Lambda^{0}=\sqrt{2}G_{F}(n_{e}-n_{\bar{e}})$
and the corresponding current $\vec{\Lambda}$. One can recognize
the similarity of the above equation to Eq. (\ref{EOM}), in which
only time-independent oscillations are considered for radially-moving
neutrinos.

We decompose $\rho$ as 
\begin{align}
\rho=\dfrac{f_{\nu_{e}}+f_{\nu_{x}}}{2}+\dfrac{f_{\nu_{e}}-f_{\nu_{x}}}{2}\begin{pmatrix}s & S\\
S^{*} & -s
\end{pmatrix}
\end{align}
with the maximum occupation numbers $f_{\nu_{e}}$ and $f_{\nu_{x}}$
just as in Eq. (\ref{decomposePhi}). Then, $s=1$ and $S=0$ corresponds
to flavor eigenstates, which are fixed points of the equations of
motion if one ignores the minor off-diagonal elements in the mass
matrix in vacuum. Linearizing Eq. (\ref{freestreaming}) in the neighborhood
of one of these fixed points, we obtain the equation for $S$ as 
\begin{align}
i(\partial_{t}+\vec{v}\cdot\boldsymbol{\nabla}_{\vec{r}})S_{\vec{v}}=v^{\mu}(\Lambda_{\mu}+\Phi_{\mu})S_{\vec{v}}-\int\dfrac{d\vec{v}'}{4\pi}v^{\mu}v_{\mu}'G_{\vec{v'}}S_{\vec{v'}},
\end{align}
where we define $G_{\vec{v}}$ and $\Phi^{\mu}$ as 
\begin{align}
G_{\vec{v}}=\sqrt{2}G_{F}\int_{0}^{\infty}\dfrac{dEE^{2}}{2\pi^{2}}\left[f_{\nu_{e}}(E,\vec{v})-f_{\bar{\nu}_{e}}(E,\vec{v})\right]\label{gv}
\end{align}
and
\begin{align}
\Phi^{\mu}\equiv & \int\dfrac{d\vec{v}}{4\pi}G_{\vec{v}}v^{\mu}.
\end{align}
Assuming the following form of solution in the local approximation,
$S_{\vec{v}}=Q_{\vec{v}}e^{-i(\Omega t-\vec{K}\cdot\vec{r})}$ we
obtain again the homogeneous integral equation for $Q_{\vec{v}}$
as follows: 
\begin{align}
v^{\mu}k_{\mu}Q_{\vec{v}}=-\int\dfrac{d\vec{v}'}{4\pi}v^{\mu}v_{\mu}'G_{\vec{v}'}Q_{\vec{v}'},\label{EigenEq2}
\end{align}
where we introduce $k^{\mu}=(\omega,\vec{k})=K^{\mu}-\Lambda^{\mu}-\Phi^{\mu}$.
Since the right hand side of this equation is expressed as $v^{\mu}a_{\mu}$
with 
\begin{align}
a^{\mu}\equiv-\int\dfrac{d\vec{v}}{4\pi}v^{\mu}G_{\vec{v}}Q_{\vec{v}},
\end{align}
we can write $Q_{\vec{v}}$ as $Q_{\vec{v}}=v^{\mu}a_{\mu}/v^{\mu}k_{\mu}$.
Substituting this back into Eq. (\ref{EigenEq2}), we obtain the equation
for $a^{\mu}$ as 
\begin{align}
\Pi^{\mu\nu}(\omega,\vec{k})a_{\nu}=0,
\end{align}
in which $\Pi^{\mu\nu}$ is given as 
\begin{align}
\Pi^{\rho\sigma} & =\eta^{\rho\sigma}+\int\dfrac{d\vec{v}}{4\pi}G_{\vec{v}}\dfrac{v^{\rho}v^{\sigma}}{\omega-\vec{v}\cdot\vec{k}}\nonumber \\
 & =\eta^{\rho\sigma}+\int_{-1}^{1}d\mu\dfrac{1}{\omega-k\mu}G^{\rho\sigma}(\mu)\label{Pi}
\end{align}
with 
\begin{align}
G^{\rho\sigma}(\mu)\equiv\int_{0}^{2\pi}\dfrac{d\phi}{4\pi}G_{\vec{v}}v^{\rho}v^{\sigma}.
\end{align}
In these equations, $\mu$ is the cosine of the zenith angle $\theta$
and $\phi$ is the azimuthal angle in the polar coordinates, which
are chosen so that the zenith should be oriented in the direction
of $\vec{k}$. Then the following condition has to be satisfied: 
\begin{align}
D(\omega,\vec{k})\equiv\det\Pi=0,\label{DR}
\end{align}
which finally gives us the dispersion relation (DR) between $\omega$
and $\vec{k}$. It should be noted that the integral in Eq. (\ref{Pi})
has a quite similar structure to those in $I_{n}$ given in Eq. (\ref{I}),
the fact that is eventually responsible for the appearance of the
spurious modes also in this approach.

\begin{figure}[tbh]
\includegraphics[width=7cm]{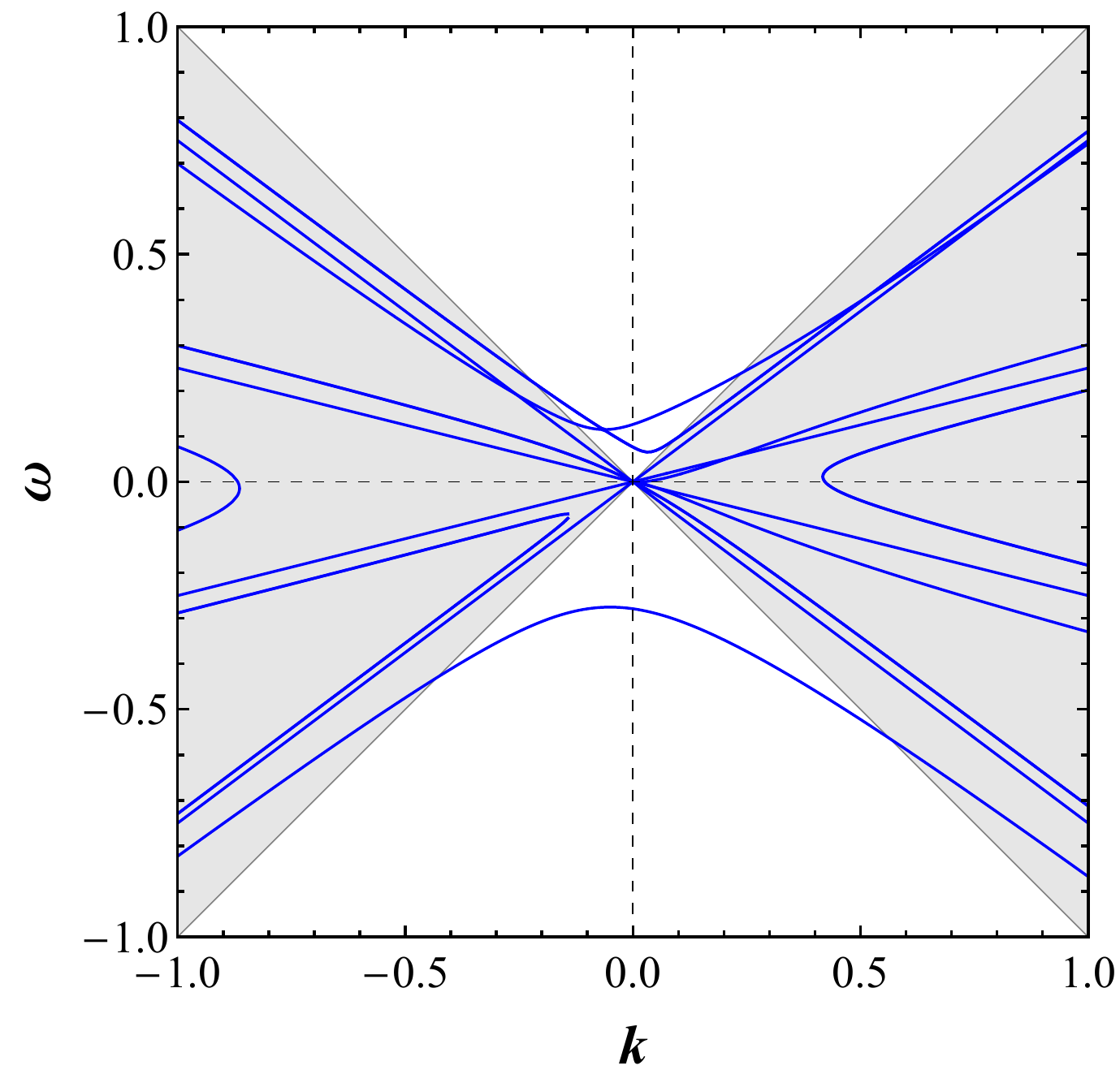} \includegraphics[width=7cm]{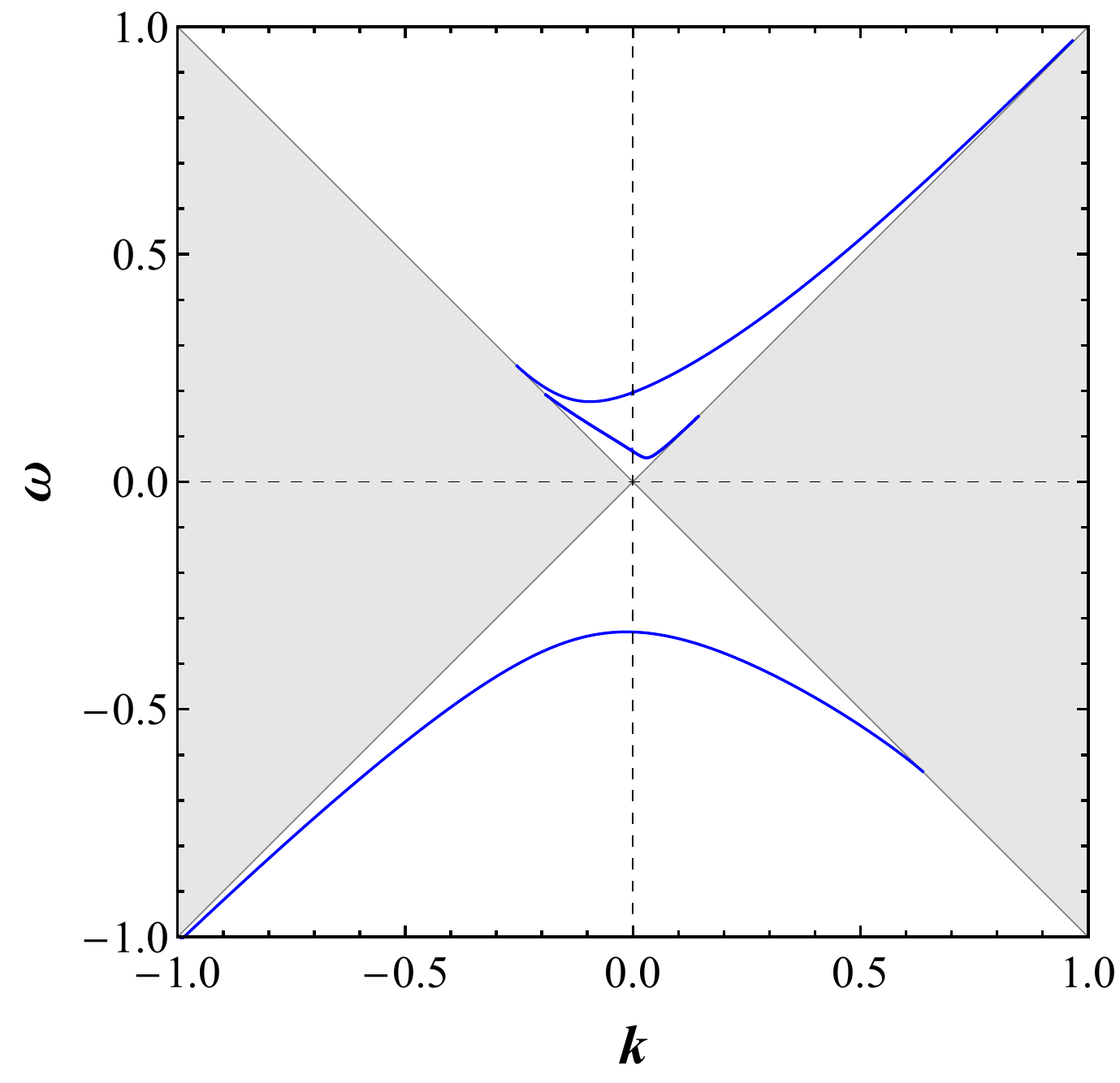}
\caption{The dispersion relation for a toy model with $G_{\vec{v}}\propto(\mu+1)\mu(\mu-0.65)$.
The top panel displays the results obtained by the discretization
of the $\mu$ integrals with 4 angular bins while in the bottom panel
we present the results of the analytical-integration method. The shaded
regions are the zone of avoidance. }
\label{dr_gtoy} 
\end{figure}

In the DR approach we first search for solutions of Eq. (\ref{DR}),
in which $\omega$ and $k$ are both real. They normally form several
branches. If there opens a gap in $\omega$ among these branches,
that is, there is no solution with real $k$ for the range of real
$\omega$, then $k$ should be complex and an instability occurs in
the spatial regime. If, on the other hand, a gap opens in $k$, an
instability in the temporal regime should occur. The DR approach is
hence very convenient to judge in which regime the instability occurs
(see also Ref. \cite{PhysRevD.96.043016}). Note, however, that we
still need to solve Eq. (\ref{DR}) somehow to obtain complex solutions
and the spurious modes also obtain if one were to solve Eq. (\ref{DR})
either to obtain $\omega$ for a given $k$ or to find $k=|\vec{k}|$
for a given $\omega$ and the direction of $\vec{k}$ by discretizing
the angular distribution $G_{\vec{v}}$. This can be inferred from
the similarity of the integral in Eq. (\ref{Pi}) to those in $I_{n}$.
Fig. \ref{dr_gtoy} shows the DR obtained by the discretization method
with 4 angular bins for a toy model with $G_{\vec{v}}\propto(\mu+1)\mu(\mu-0.65)$.
Note that this distribution is axisymmetric with respect to $\vec{k}$.
Such distributions are expected only for $\vec{k}$ parallel to the
radial direction in spherically symmetric background. Otherwise, i.e.,
either for a non-radial $\vec{k}$ or in non-spherical background,
$G_{\vec{v}}$ should have a $\phi$-dependence in general. Even in
that case the present method works just as well \cite{InPreparation1}.
In this figure the negative $k$ corresponds to the mode going in
the opposite direction and $\mu$ is replaced with $-\mu$ in $G_{\vec{v}}$.
In this toy model, DR consists of many branches, only three of which
are true modes and the rest are spurious. It is important that all
spurious branches lie in the so-called zone of avoidance \cite{PhysRevLett.118.021101},
in which $\omega=ck$ is satisfied for $|c|\leq1$ and the integral
in Eq. (\ref{Pi}) diverges. This fact also explains why the spurious
modes occur in the discretization method. If the integral in Eq. (\ref{Pi})
is conducted approximately by the discretization, then the integrand
is evaluated only at a finite number of points and the integral does
not diverge even for combinations of $\omega$ and $k$ in the zone
of avoidance unless one of the sampling points accidentally coincides
with $k$/$\omega$. The zone of avoidance is hence useful to judge
which branch is spurious at a glance. 

\begin{figure}[tbh]
\includegraphics[width=7cm]{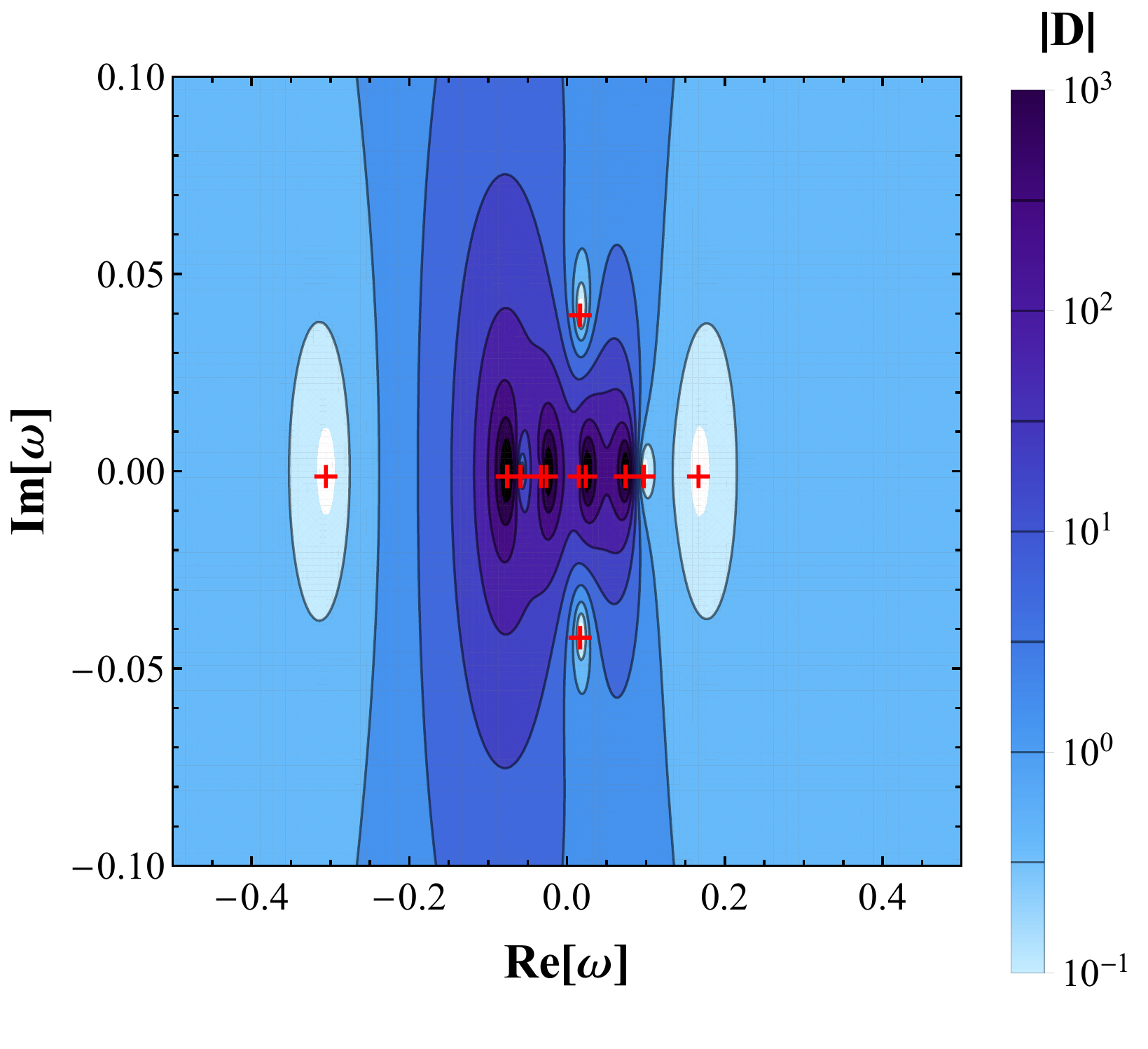} \includegraphics[width=7cm]{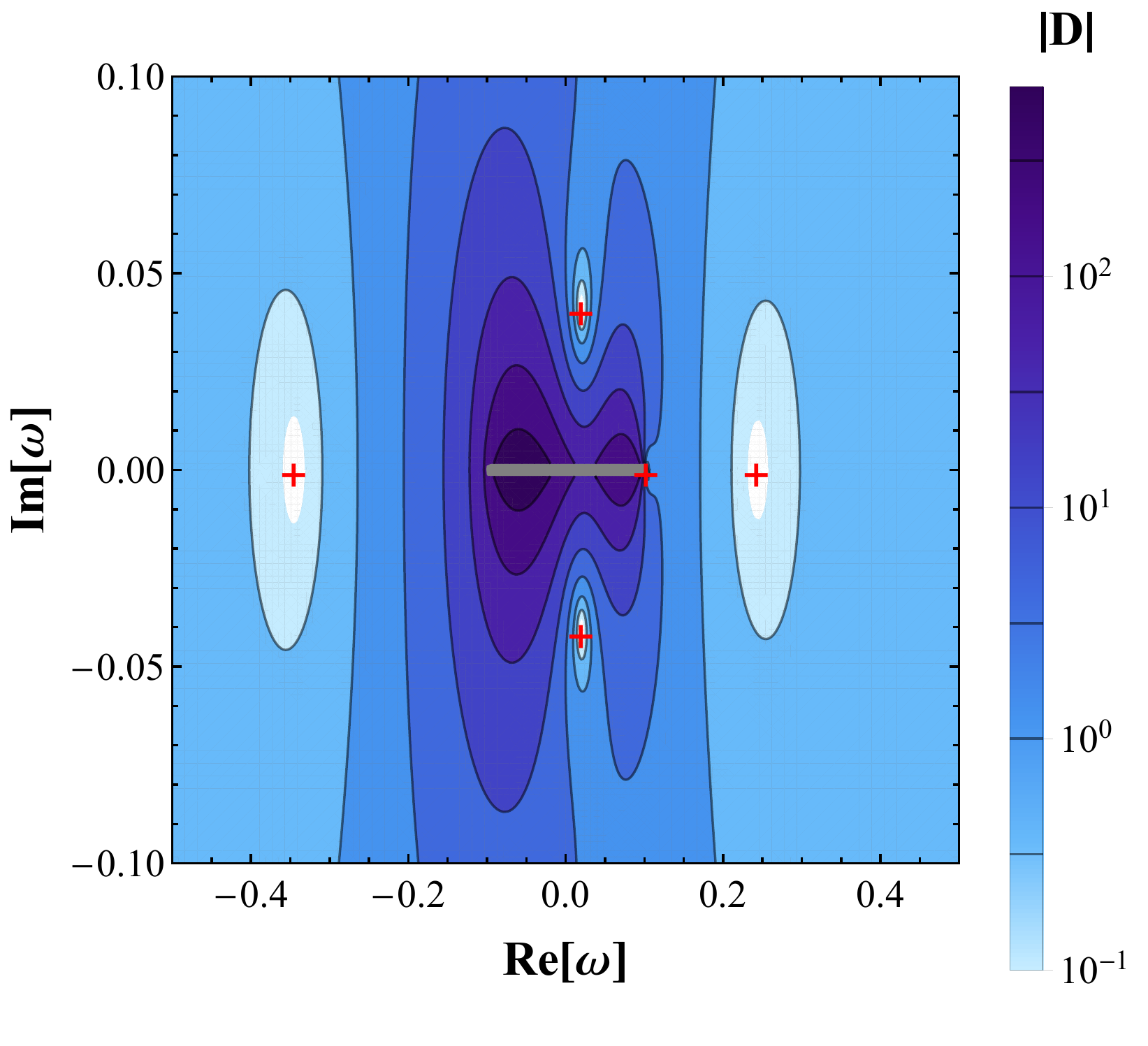}
\caption{The Absolute values of $D(\omega,\vec{k})$ in the complex plane of
$\omega$ for $k=0.1$. The angular distribution $G_{\vec{v}}$ is
the same as in Fig. \ref{dr_gtoy}. The top panel displays the results
obtained by the discretization of the $\mu$ integrals with 4 angular
bins while in the bottom panel we present the results of the analytical-integration
method. Red pluses mark the positions of the zero points of $D$ in
each method. The gray line in the bottom panel indicates the branch
cut of the Riemann surface obtained in the analytical-integration
method.}
\label{contour_gtoy} 
\end{figure}

The complex solutions are more involved. In fact, judging from the
gap in $\omega$ for the three true branches, one may think that there
will be instabilities only in the spatial regime. This is not true,
however. In fact, we show in Fig. \ref{contour_gtoy} the absolute
values of $D(\omega,\vec{k})$ in the complex $\omega$ plain for
$k=0.1$. In the upper panel, where the results of the discretization
approximation are shown, there are many real spurious modes and only
outer three modes ($\omega\approx-0.3$, 0.1 and 0.15) are true modes
as mentioned above (see also Fig. \ref{dr_gtoy}). What is more important
here is the fact that there exist complex true modes also (see the
lower panel of Fig. \ref{contour_gtoy}), which cannot be recognized
from DR alone. We hence have to search for complex solutions of Eq.
(\ref{DR}) somehow and again face the same problem of the spurious
modes. 

\subsection{Analytical-Integration Approach}

\begin{figure}[tbh]
\includegraphics[width=7cm]{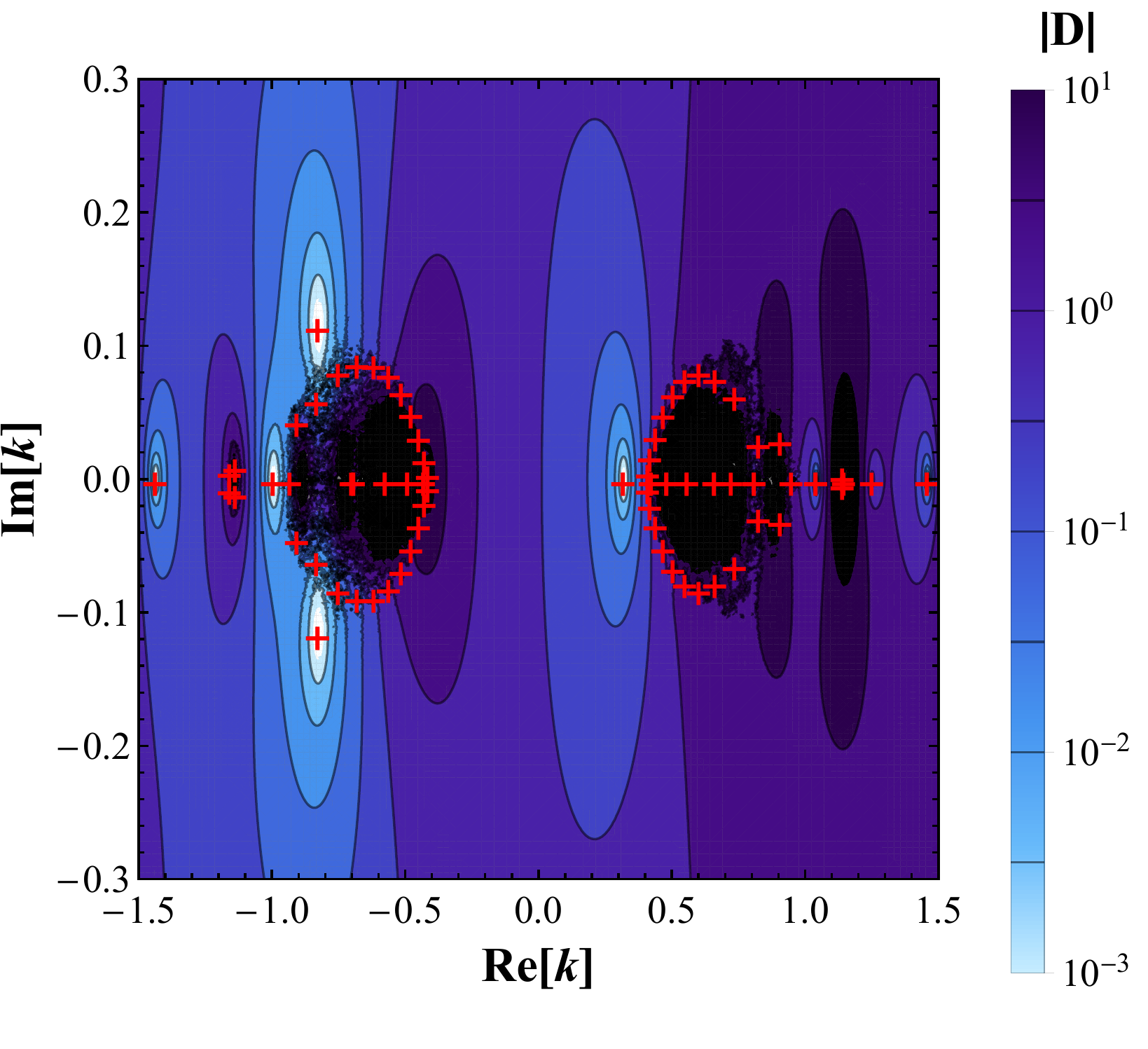} \includegraphics[width=7cm]{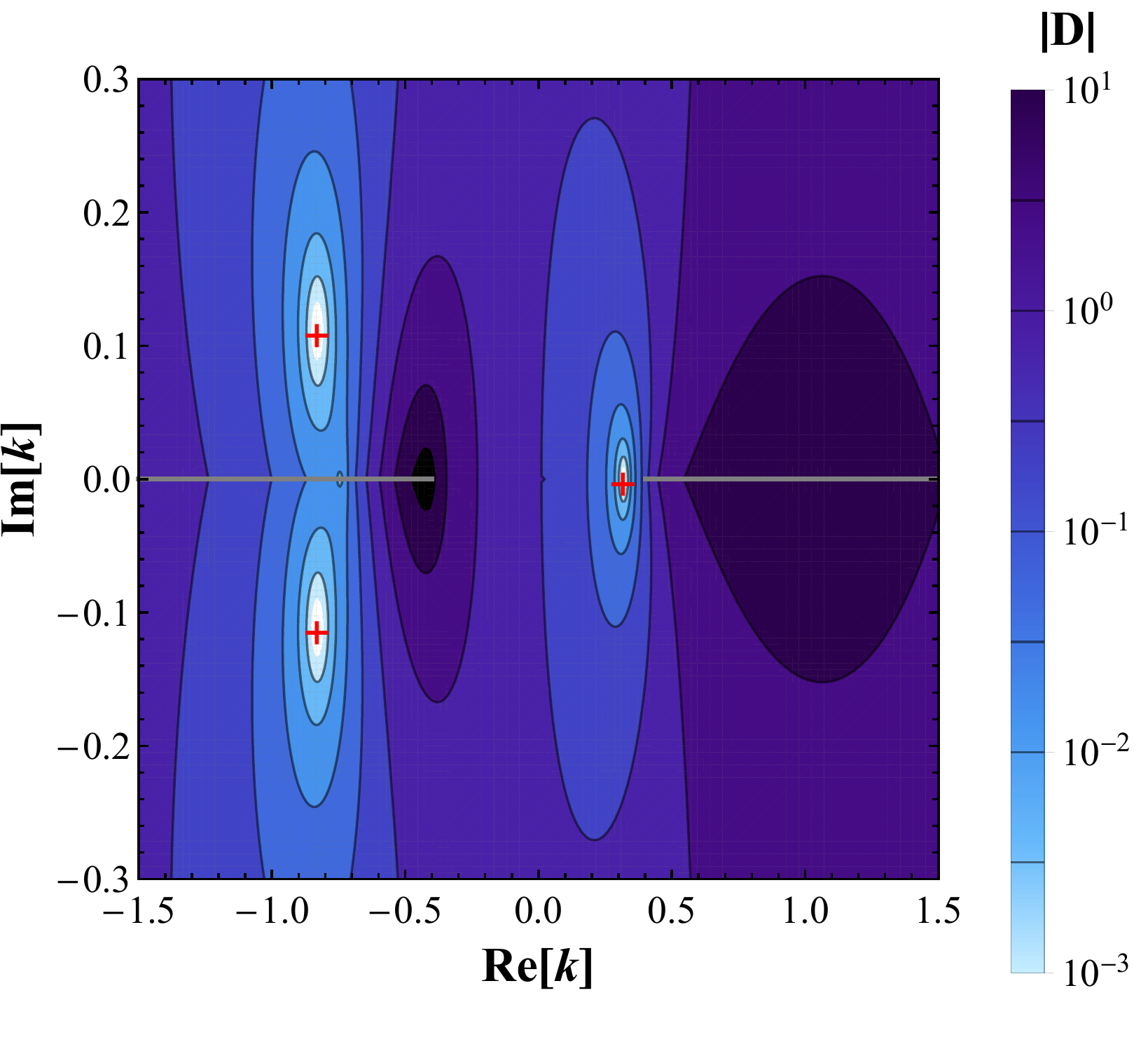}
\caption{The Absolute values of $D(\omega,\vec{k})$ in the complex plane of
$k$ for $\omega=0.4$ for a toy model with $G_{\vec{v}}\propto-(\mu+0.5)(\mu-1)$.
The top panel displays the results obtained by the discretization
of the $\mu$ integrals with 20 angular bins while in the bottom panel
we present the results of the analytical-integration method. Red pluses
mark the positions of the zero points of $D$ in each method. The
gray lines in the bottom panel indicate the branch cuts of the Riemann
surface obtained in the analytical-integration method.}
\label{contour_gtoy2} 
\end{figure}
\begin{figure}[htb]
\includegraphics[width=7cm]{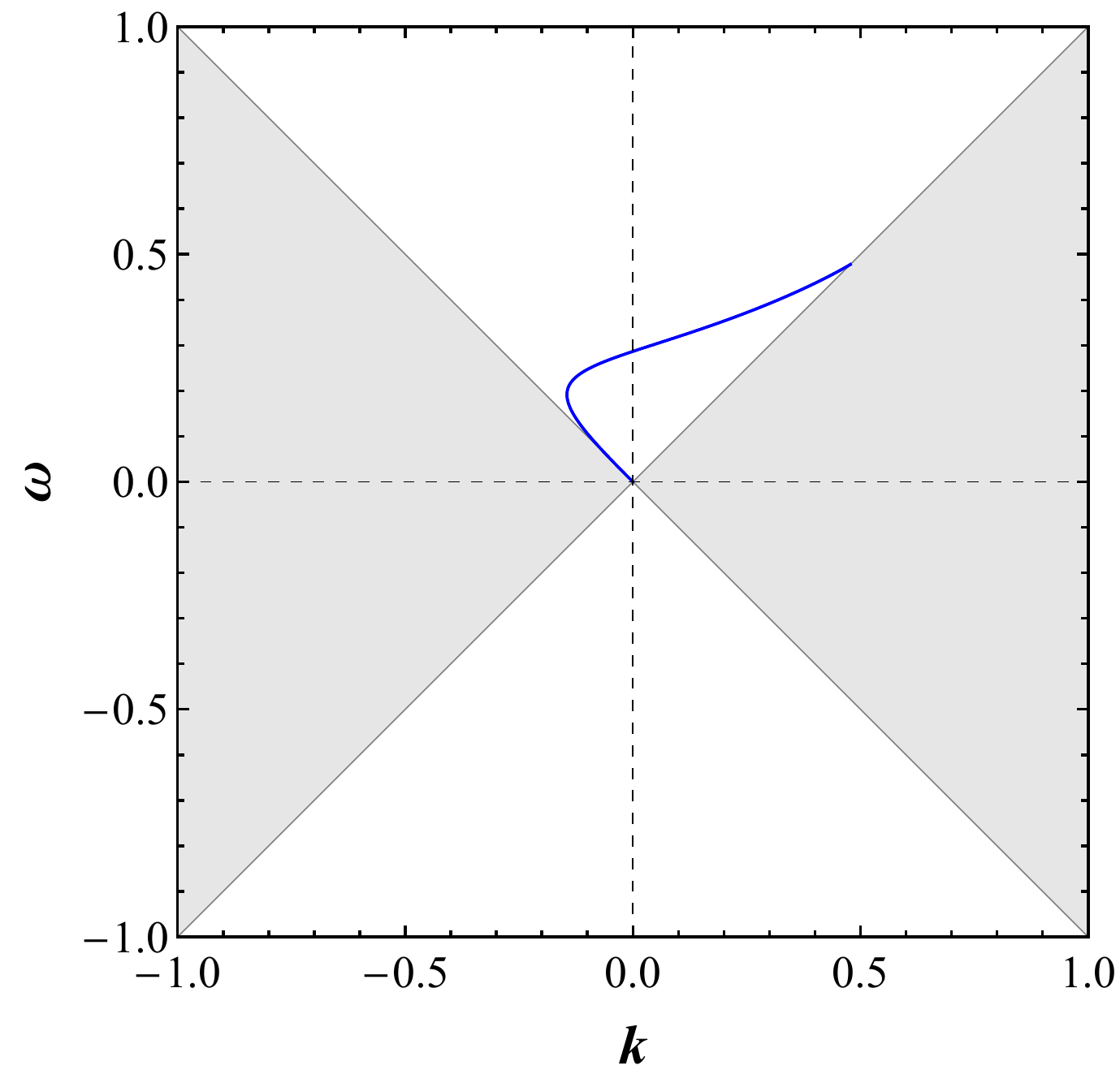}\caption{The dispersion relation obtained in the analytical-integration method
for the same toy model as in Fig. \ref{contour_gtoy2}. The shaded
regions are the zone of avoidance. }
\label{dr_gtoy2} 
\end{figure}
Although there was no complex spurious modes in the previous model,
this is not the case in general. In fact, we show in the upper panel
of Fig. \ref{contour_gtoy2} the complex spurious solutions obtained
for the case of $G_{\vec{v}}\propto-(\mu+0.5)(\mu-1)$ in the discretization
method with 20 angular bins. The corresponding DR is presented in
Fig. \ref{dr_gtoy2}. These spurious modes can be avoided by approximating
$G(\mu)$ polynomially again. Then we can evaluate the remaining integrals
with respect to the energy $E$ and the azimuthal angle $\phi$ just
by discretization, since they do not yield poles of $\omega$ or $k$. 

The concrete procedure is the following: each component of $G^{\rho\sigma}(\mu)$
is first approximated as 
\begin{align}
G^{\rho\sigma}(\mu)\simeq\sum_{i=0}^{d}g_{i}^{\rho\sigma}\mu^{i};
\end{align}
then the integrals in $\Pi$ are performed analytically as 

\begin{align}
 & \int_{-1}^{1}d\mu\dfrac{1}{\omega-k\mu}G^{\rho\sigma}(\mu)\nonumber \\
 & \simeq\sum_{i=0}^{d}g_{i}^{\rho\sigma}\int_{-1}^{1}d\mu\dfrac{\mu^{i}}{\omega-k\mu}\nonumber \\
 & =\dfrac{1}{k}\sum_{i=0}^{d}g_{i}^{\rho\sigma}\biggl[\left(\dfrac{\omega}{k}\right)^{i}\ln\dfrac{\omega+k}{\omega-k}-\sum_{m=1}^{\ceiling{\frac{i}{2}}}\dfrac{2}{2m-1}\left(\dfrac{\omega}{k}\right)^{i-2m+1}\biggr];
\end{align}
finally the determinant of $\Pi$ is calculated and its zero points
are looked for either in the complex plane of $\omega$ for a given
$k$ or in the complex plane of $k$ for a given $\omega$. Note that
in both cases $D(\omega,\vec{k})$ has a branch cut in the Riemann
surface of $\omega$ or $k$ on the parts of the real axis that satisfy
$|\omega|\leq|k|$ for each $k$ or $\omega$. We show in the bottom
panel of Fig. \ref{contour_gtoy} the absolute values of $D(\omega,\vec{k})$
in the complex $\omega$ plane obtained in this way. The value of
$k$ is set to 0.1. It is evident that the spurious modes are all
gone and only the true modes, three real and two complex ones, are
remaining. In the bottom panel of Fig. \ref{contour_gtoy2}, in which
an example that has complex spurious modes is exhibited, we demonstrate
that our method can also eradicate these complex spurious modes successfully.
The reason should be now clear: the essential feature in the Riemann
surface is maintained in the analytical-integration approach.

\section{Applications to Simulation Data}

As mentioned earlier, we have in mind the applications of our method
to more realistic data provided by numerical simulations. Then the
neutrino distribution functions available are discrete in energy and
angles from the beginning. It is hence the purpose of this section
to demonstrate that such discrete data can be handled without any
difficulty by our method. 
\begin{figure}[htb]
\includegraphics[width=7cm]{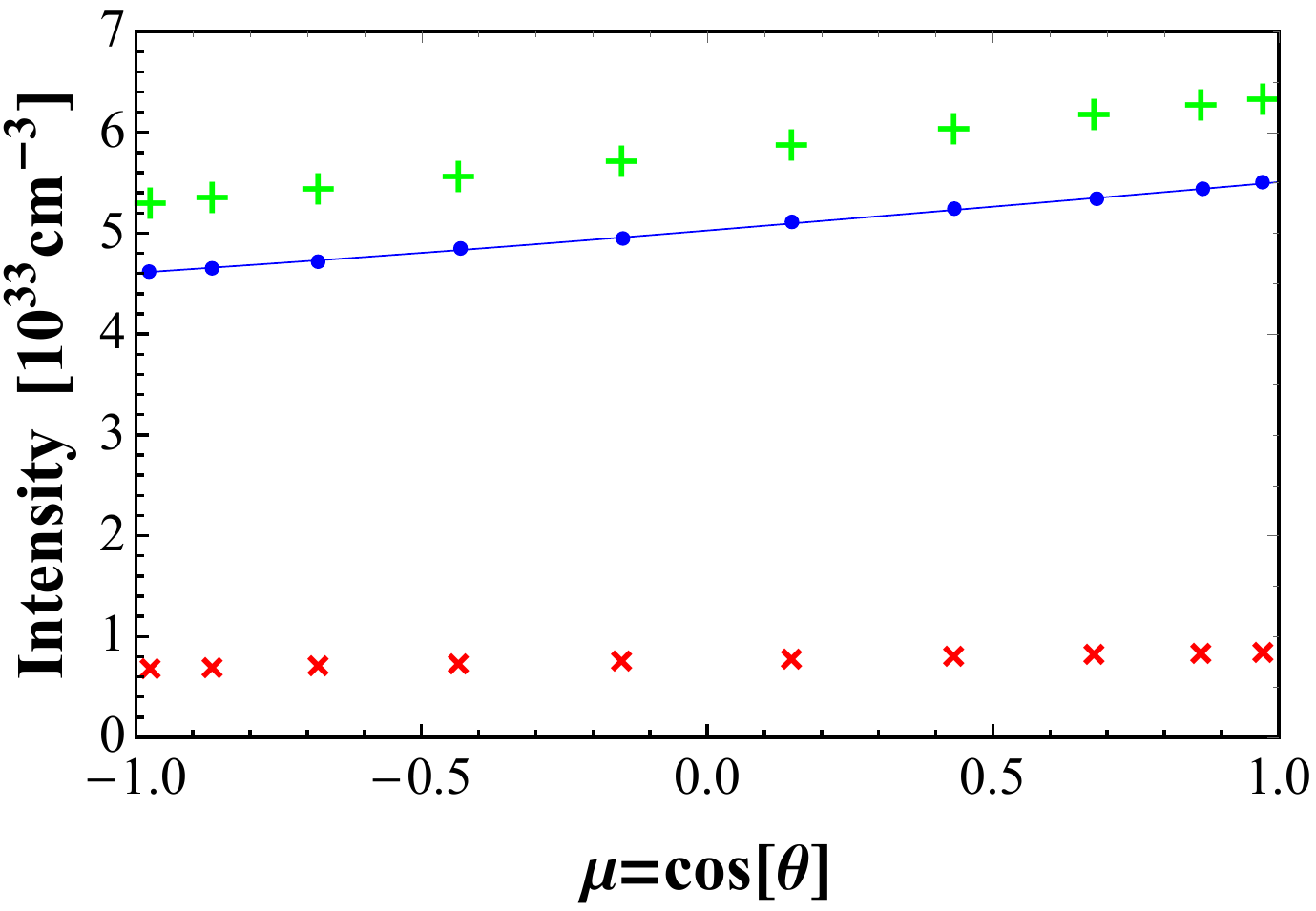} \caption{The angular distributions of $\nu_{e}$ (green pluses) and $\bar{\nu}_{e}$
(red crosses) at the time of 200 ms post bounce and at the radius
of 40 km obtained in a SN simulation. The blue points are the difference
between them and the blue solid line is the polynomial approximation
to it. }
\label{g-mu_sn} 
\end{figure}

Here we employ the neutrino distributions obtained in a spherically
symmetric simulation of core collapse supernovae \cite{PrivateCommunication}.
The progenitor is a non-rotating massive star of $11.2M_{\odot}$
\cite{RevModPhys.74.1015}. The dynamics of core collapse, bounce
and shock stagnation is computed with a Boltzmann-radiation-hydrodynamics
code, for details of which we refer readers to \cite{0067-0049-214-2-16,0067-0049-229-2-42}:
it solves hydrodynamics equations and Boltzmann equations for neutrino
transport simultaneously; Newtonian self-gravity is implemented; a
realistic equation of state based on the relativistic mean field theory
for uniform nuclear matter is adopted \cite{PrivateCommunication}.
This model fails to produce an explosion as is normally the case in
spherically symmetric simulations. The neutrino distribution functions
we use here are taken from the snapshot at the post-bounce time of
200ms. The radial position is $r=40\mathrm{km}$. 

\begin{figure}[htb]
\includegraphics[width=7cm]{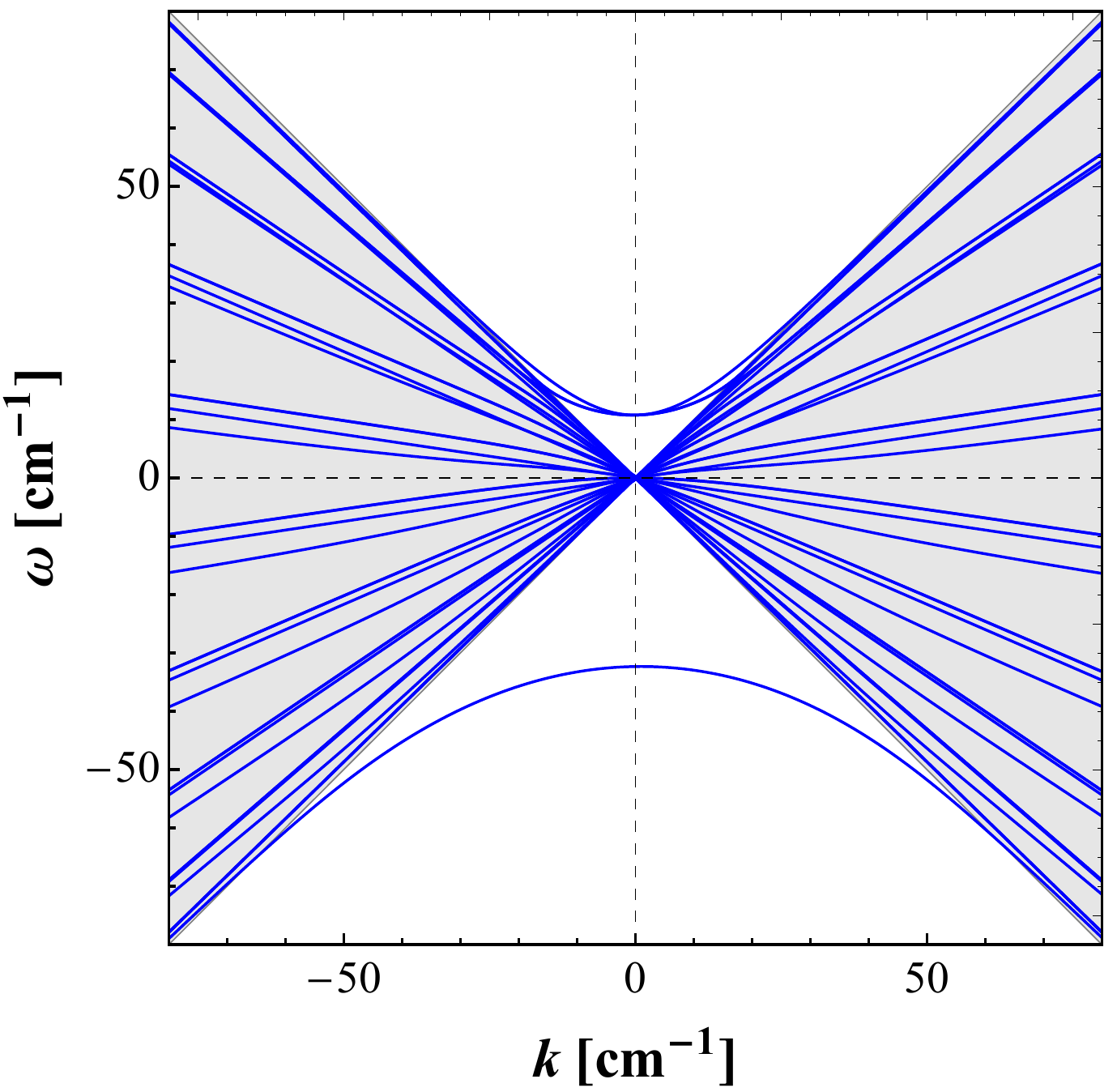} \includegraphics[width=7cm]{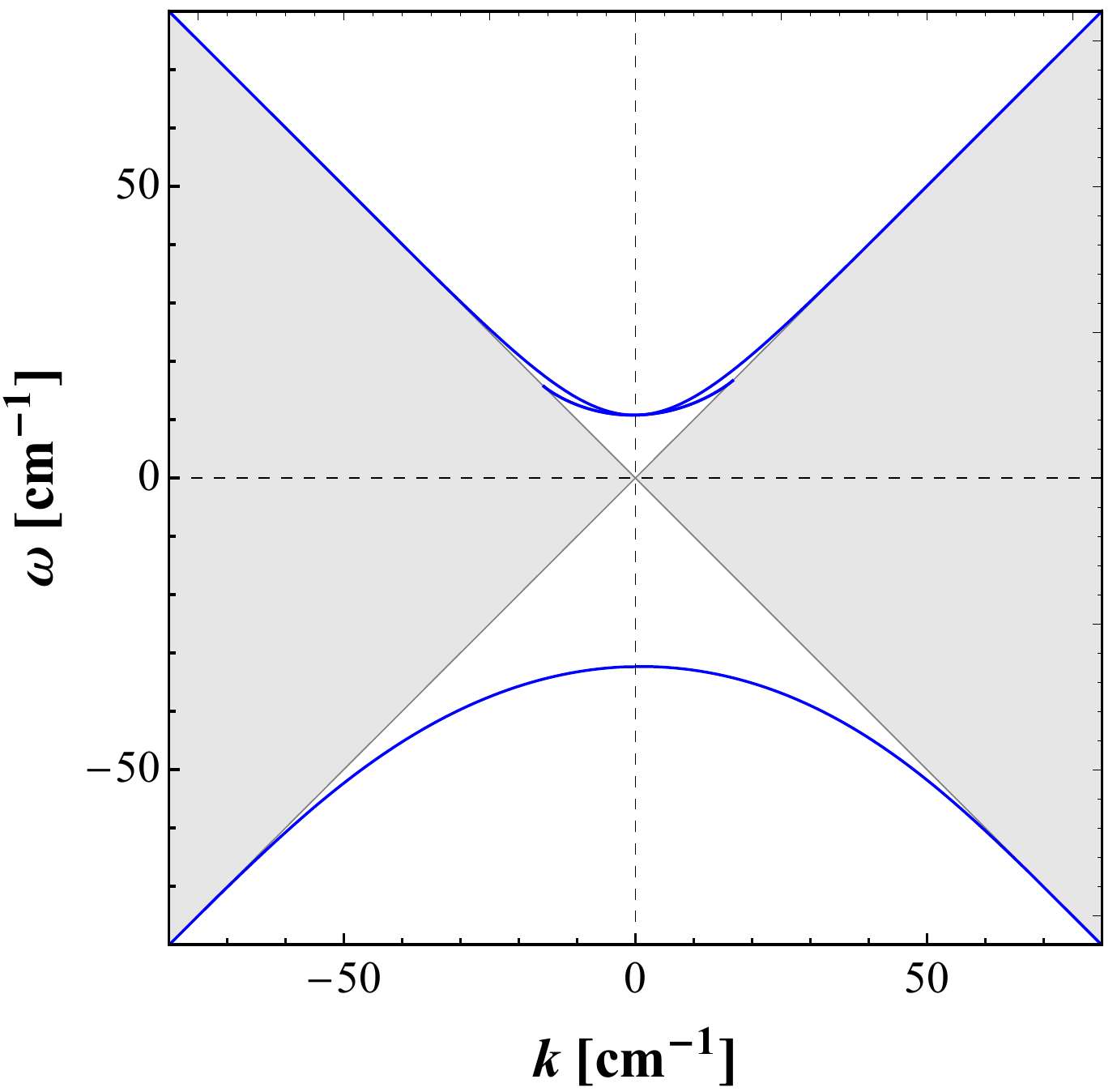}
\caption{Dispersion relations for the angular distribution in Fig. \ref{g-mu_sn}.
The top panel displays the results obtained by the discretization
approximation while in the bottom panel we present the results of
the analytical-integration method. The shaded regions are the zone
of avoidance. }
\label{dr_sn} 
\end{figure}

\begin{figure}[htb]
\includegraphics[width=7cm]{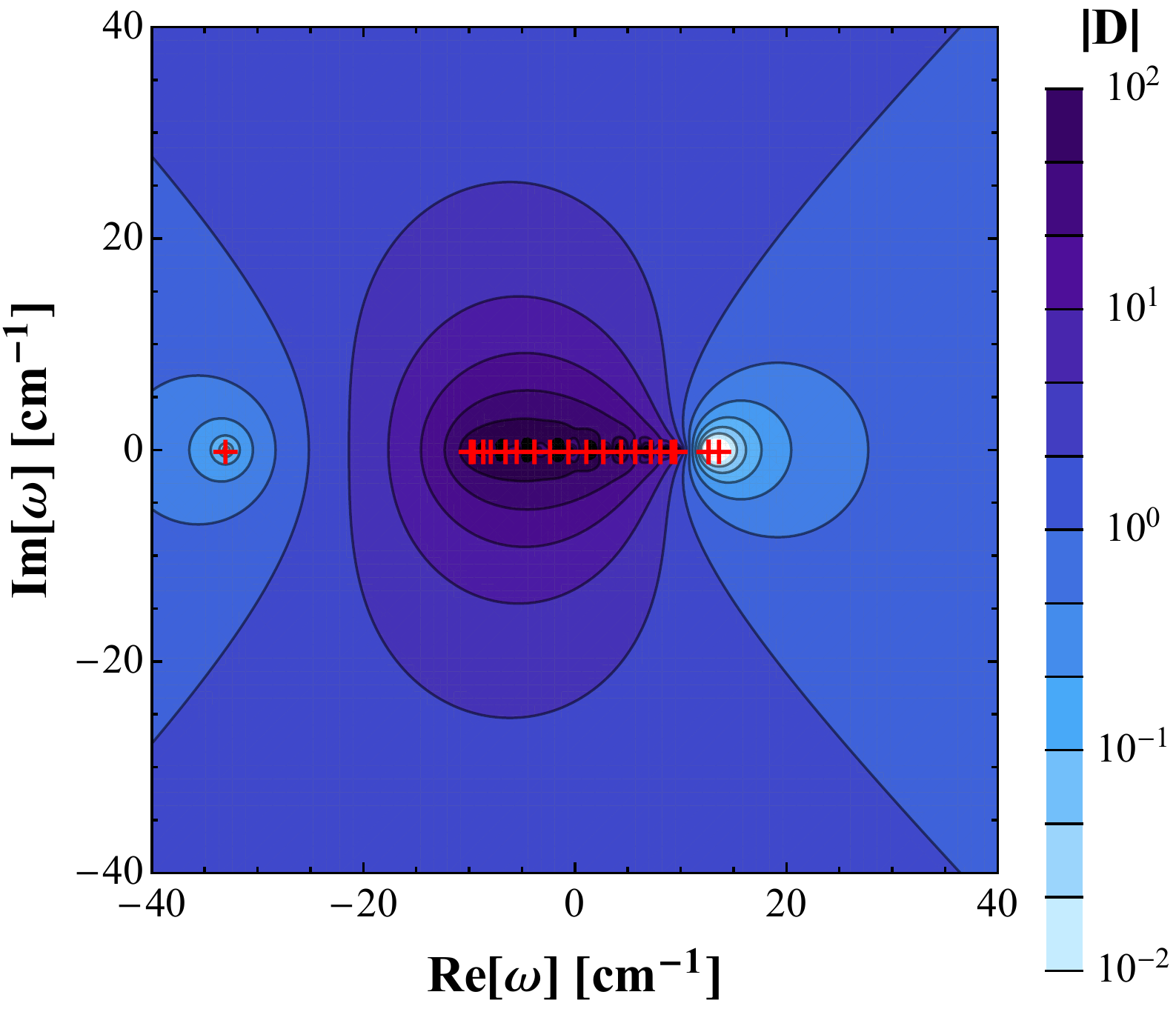} \includegraphics[width=7cm]{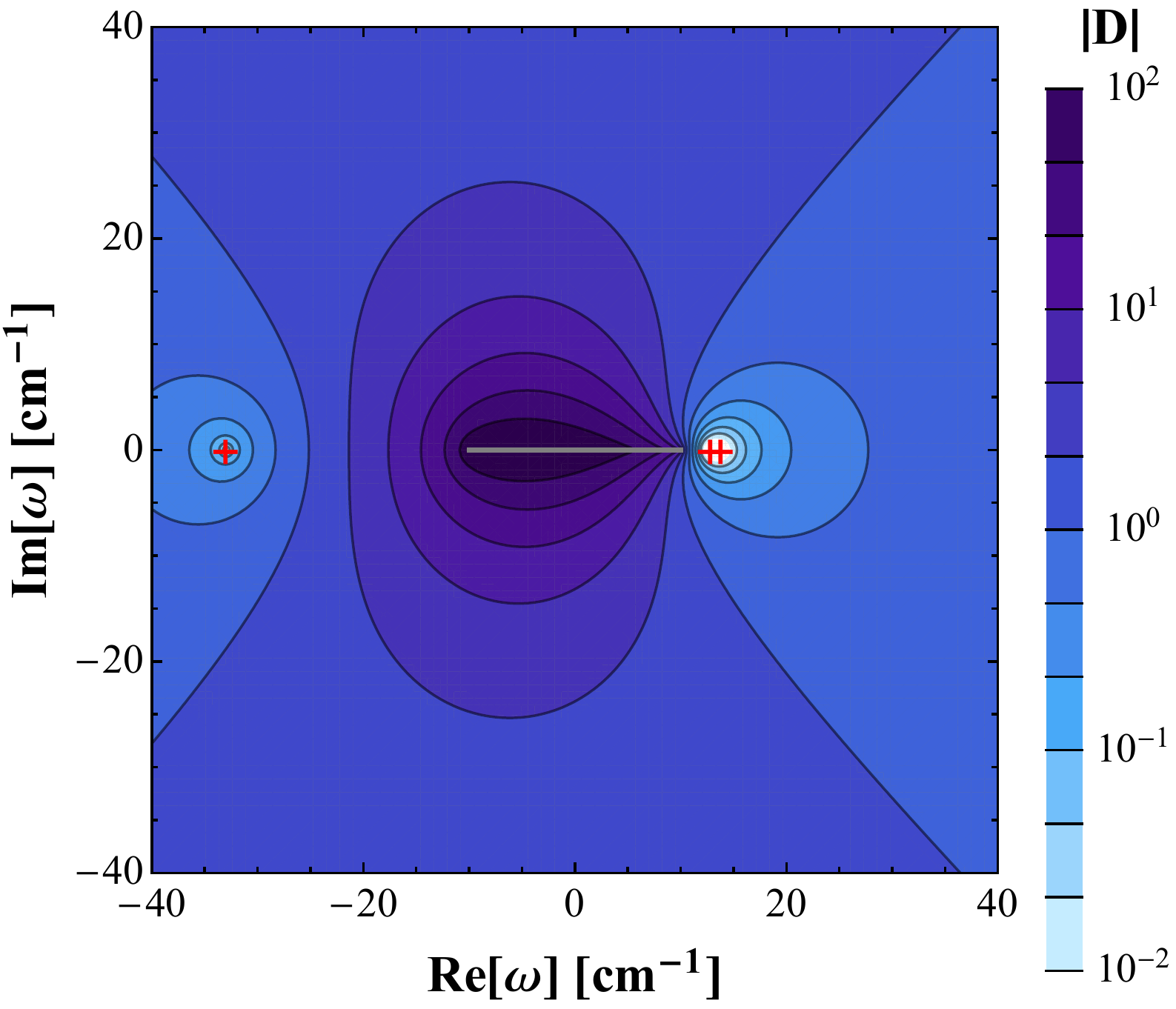}
\caption{The Absolute values of $D(\omega,\vec{k})$ in the complex plane of
$\omega$ for $k=10\mathrm{cm}^{-1}$ for the angular distribution
given in Fig. \ref{g-mu_sn}. The top panel displays the results obtained
by the discretization approximation while in the bottom panel we present
the results of the analytical-integration method. Red pluses mark
the positions of the zero points of $D$ in each method. The gray
line in the bottom panel indicates the branch cut of the Riemann surface
obtained in the analytical-integration method. }
\label{contour_sn} 
\end{figure}

\begin{figure}[htb]
\includegraphics[width=7cm]{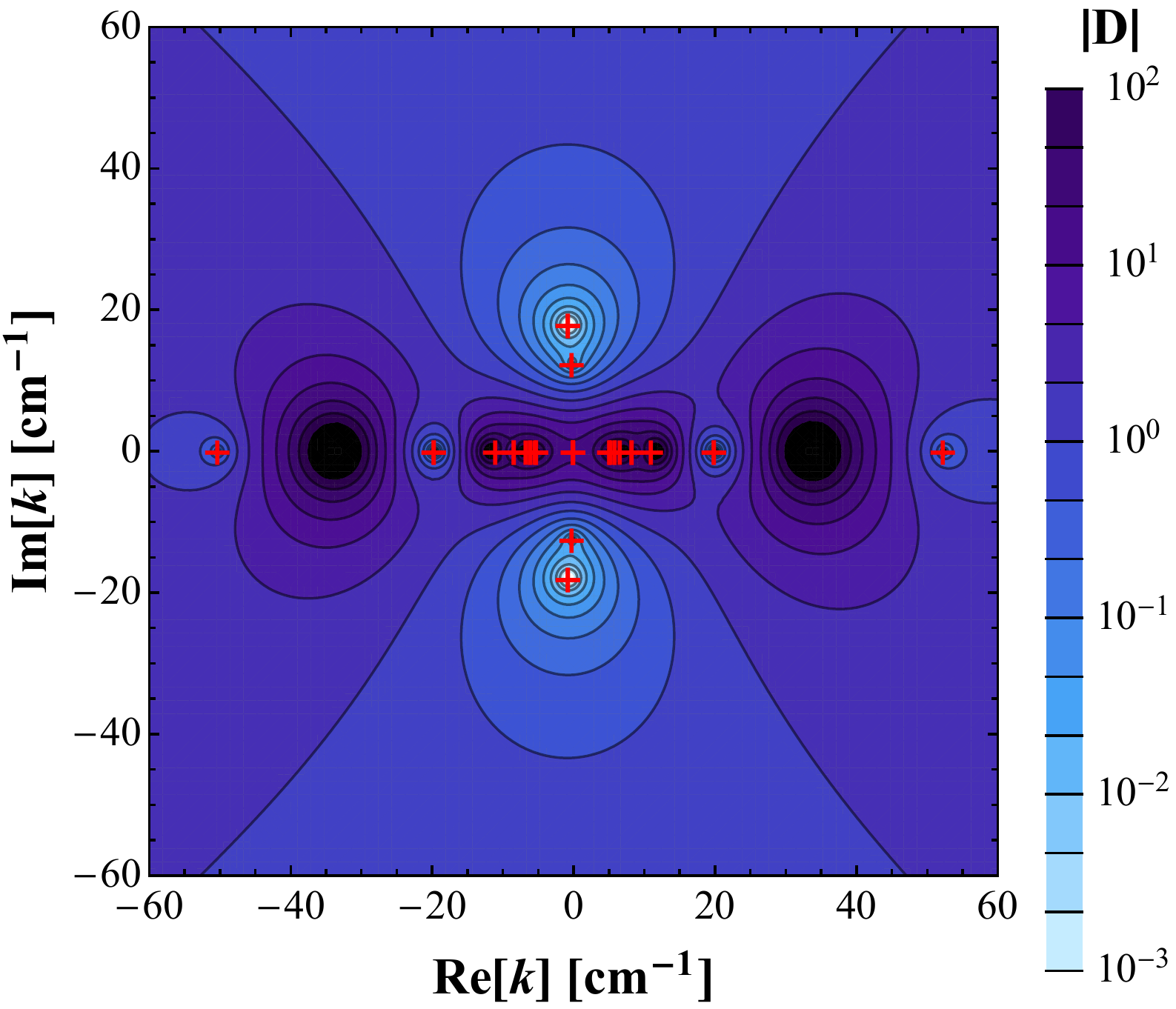} \includegraphics[width=7cm]{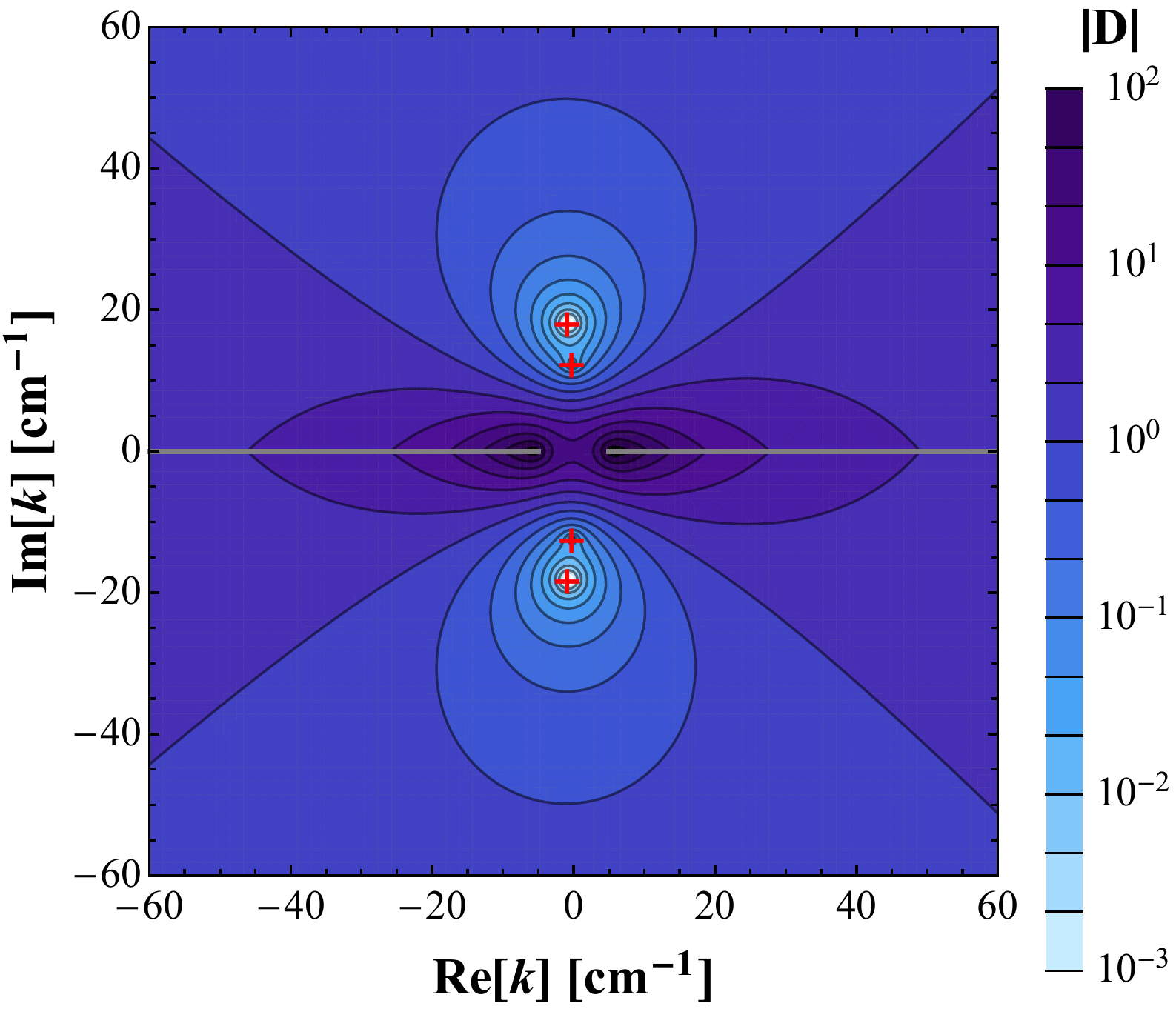}
\caption{The Absolute values of $D(\omega,\vec{k})$ in the complex plane of
$k$ for $\omega=5\mathrm{cm}^{-1}$ for the angular distribution
given in Fig. \ref{g-mu_sn}. The top panel displays the results obtained
by the discretization approximation while in the bottom panel we present
the results of the analytical-integration method. Red pluses mark
the positions of the zero points of $D$ in each method. The gray
lines in the bottom panel indicate the branch cuts of the Riemann
surface obtained in the analytical-integration method. }
\label{contour_sn_k} 
\end{figure}

The angular distributions of neutrinos that we employ are shown in
Fig. \ref{g-mu_sn}. We note that it is the difference between the
intensities of $\nu_{e}$ and $\bar{\nu_{e}}$ that is most important
for the collective flavor oscillation as is understood from Eq. (\ref{gv}).
Figures \ref{dr_sn} and \ref{contour_sn} show the DR and absolute
values of $D$, respectively, in which the upper panels are obtained
by discretization whereas the lower ones are the results of the analytical-integration
method. From the former it is apparent that many spurious modes appear
when we evaluate the integral in Eq. (\ref{Pi}) by discretization.
In sharp contrast the analytical-integration method generate none
of them. In this case, the gap opens in $\omega$, implying a possible
instability in the spatial regime (see Fig. \ref{contour_sn_k}).
It is also confirmed with no difficulty in our method that there is
no unstable mode in the temporal regime for, e.g., $k=10\mathrm{cm}$
as demonstrated in the bottom panel of Fig. \ref{contour_sn}. 

\section{Conclusions}

The appearance of the spurious modes is a vexing problem in the analysis
of collective flavor oscillations of neutrino in the supernova core.
They emerge even in the linear analysis more often than not when one
solves the integro-differential equations by the ordinary discretization.
In principle, they can be distinguished from the true modes, since
the spurious ones do not converge to real solutions as the number
of bins employed for the discretization is increased. Such procedures
are inefficient, however, if one were to analyze numerical data obtained
in realistic simulations systematically and, if possible, in real
time. In this paper, we have analyzed in detail why the spurious modes
appear in the first place and have proposed a simple method to avoid
them from the beginning in the local linear analysis. 

We have found that the ultimate source of the spurious modes is a
generation of pole singularities in the approximate angular integrations
by the discretization, in which the angular distribution of neutrino
is expressed as a superposition of delta functions. The exact integration
would produce two branching points and a cut in between instead. It
is hence reasonable to consider that the spurious modes will not appear
if one retain the singularity structure in the approximation. The
easiest way to do this may be to approximate the angular distribution
with polynomials and perform the angular integration analytically.
We have demonstrated for some toy models that the idea really works
as expected. We have started with the time-independent mode propagating
in the radial direction under the background of matter distributed
spherically in space and monochromatic neutrinos emitted semi-isotropically
from the neutrino sphere. Note that polynomials are not the unique
option but any base functions will work equally well as long as they
do not change the singularity structure. We need to strike a balance
then between accuracy of the approximation and easiness of the integration.
We have also shown that the condition may be relaxed a bit: the single
branch cut may be replaced by a union of sub-branch cuts that are
produced, for example, in the piecewise constant approximation. In
this case the pole singularities are replaced by branching-point singularities
but no spurious mode is produced. This finding is important in applying
the method to numerical data, which are normally provided only on
a set of discrete grid points. 

We have then considered a multi-energy case. We have observed that
it is sufficient to apply the polynomial approximation or the piecewise
constant approximation only to the angular integral. The remaining
integral with respect to the energy can be done simply by discretization.
This is understood from the fact that no pole is produced in these
methods.

Our method can be also applied to the dispersion relation approach,
which was proposed more recently and treats the instabilities both
in the spatial and temporal regimes on the equal basis. We have shown
that only the spurious modes with real $\omega$ and $k$ exist in
the zone of avoidance and argued that the existence of such modes
is closely related with the reason why the spurious modes appear in
the first place. The spurious modes are not restricted to the zone
of avoidance, though. We have demonstrated indeed that they can occur
in the temporal regime even when a gap is opening in $\omega$ and
a spatial instability is expected. The opposite is also possible.
More importantly, we have confirmed that these spurious modes are
all eliminated by our method again. This is mainly because the integro-differential
equations in the dispersion relation approach have essentially the
same structure as those for the time-independent modes. The point
is again that the approximate evaluation of the integral should not
produce pole singularities when the branching singularities are expected.
We have finally applied our method to numerical data obtained in a
realistic supernova simulation in spherical symmetry in space. We
have observed that there occurs no spurious mode. Although only radially-propagating
modes have been considered in this calculation, we believe that this
is sufficient to demonstrate that our method works just as well for
realistic data.

Modes with non-radial $\vec{k}$ can be also treated in our method.
In that case, the azimuthal dependence of the angular distribution
needs to be treated appropriately. Since we measure the zenith and
azimuth angles not from the radial direction but from the direction
of $\vec{k}$ in our method, the angular distribution becomes $\phi$-dependent
even in the spherically symmetric background, in which it is axisymmetric
with respect to the local radial direction. We can deal with this
by spherical-harmonics expansions and rotations of coordinates. The
details will be described in our forthcoming paper \cite{InPreparation1},
in which we will conduct linear analysis in more general settings
with the method proposed here. Note that the angular distribution
function of neutrino is no longer axisymmetric if the background is
not spherically symmetric, which is believed to be the case in the
supernova core owing to hydrodynamical instabilities \cite{Ann.Rev.Nucl.Part.Sci.66.341};
then we need to handle non-trivial $\phi$-dependence even for the
radially-propagating modes. 

In this paper, we have assumed the local approximation, which is valid
in the short-wavelength limit. In general, however, we need to take
into account the global background distribution. Then the eigen modes
cannot be given by exponential functions any longer. It remains to
be studied if our method can be extended to this global linear analysis.
If the answer is affirmative, our method may be further applied to
the analysis of the original nonlinear equations for the collective
flavor oscillations. It will be also interesting to see how the inclusion
of collision terms in linear analysis modifies the whole picture if
the collective oscillations are expected to occur near the neutrino
sphere in the linear analysis neglecting them. We are currently conducting
linear analysis with the present method for numerical data, which
have become available very recently from realistic radiation-hydrodynamics
simulations of CCSNe under axisymmetry in space, in which the Boltzmann
equations were solved directly for neutrino transfer \cite{arXiv1702.01752}.
The results will be reported elsewhere \cite{InPreparation2}.
\begin{acknowledgments}
We are grateful to Hiroki Nagakura for providing us with the data
of neutrino distributions obtained in his SN simulation. This work
is partially supported by Grant-in-Aid for Scientific Research from
the Ministry of Education, Culture, Sports, Science and Technology
of Japan (16H03986). 
\end{acknowledgments}

\providecommand{\noopsort}[1]{}\providecommand{\singleletter}[1]{#1}%

\end{document}